\documentclass[aps,prb,twocolumn,showpacs]{revtex4-1}
\pdfoutput=1
\usepackage{graphicx,epsf}
\usepackage{amsmath}
\usepackage{amssymb}
\usepackage{color}
\usepackage{esint}
\usepackage{wasysym}

\newcommand{\bB}{{\bf B}}
\newcommand{\bhat}{\hat{B}}

\newcommand{\bS}{{\bf S}}
\newcommand{\bR}{{\bf R}}
\newcommand{\bq}{{\bf q}}

\newcommand{\bp}{{\bf p}}
\newcommand{\bk}{{\bf k}}

\newcommand{\taub}{\mbox{\boldmath $\tau $}}
\newcommand{\deltab}{\mbox{\boldmath $\delta $}}
\newcommand{\gm}{\mathcal{G}}

\begin{document}

\title{Flat-band ferromagnetism in a topological Hubbard model}

\author{R. L. Doretto}
\affiliation{Instituto de F\'isica Gleb Wataghin,
                  Universidade Estadual de Campinas,
                  13083-859 Campinas, SP, Brazil}
\author{M. O. Goerbig}
\affiliation{Laboratoire de Physique des Solides,
                  CNRS UMR 8502,
                  Universit\'e Paris-Sud,
                  F-91405 Orsay Cedex, France}

\date{\today}

\begin{abstract}
We study the flat-band ferromagnetic phase of a topological Hubbard
model within a bosonization formalism and, in particular, determine
the spin-wave excitation spectrum. 
We consider a square lattice Hubbard model at $1/4$-filling whose 
free-electron term is the $\pi$-flux model with
topologically nontrivial and nearly flat energy bands.      
The electron spin is introduced such that the model either explicitly
breaks time-reversal symmetry (correlated flat-band Chern insulator)
or is invariant under time-reversal symmetry (correlated flat-band
Z$_2$ topological insulator).  
We generalize for flat-band Chern and topological insulators the
bosonization formalism [Phys. Rev. B {\bf 71}, 045339 (2005)]
previously developed for the two-dimensional electron gas in a
uniform and perpendicular magnetic field at filling factor $\nu=1$. 
We show that, within the bosonization scheme, the topological Hubbard
model is mapped into an effective interacting boson model. 
We consider the boson model at the harmonic approximation and show
that, for the correlated Chern insulator, the spin-wave excitation
spectrum is gapless while, for the correlated topological insulator,
gapped. We briefly comment on the possible effects of the boson-boson
(spin-wave--spin-wave) coupling. 
\end{abstract}
\pacs{71.10.Fd, 73.43.Cd, 73.43.Lp}

\maketitle

%71.10.Fd	Lattice fermion models (Hubbard model, etc.)
%
%73.43.Cd	Theory and modeling: in QHE
%
%73.43.Lp	Collective excitations: in QHE

%%%%%%%%%%%%%%%%%%%%%%%%%%%%%%%%%%%%%%%%%%%%%%%%%%%%%%%%%%%%%%%%%%%%%%%%%%%%%%%%%%%%%
\section{Introduction}
\label{sec:intro}

Electronic bands with non-zero Chern numbers are at the origin of a
large variety of topological phenomena in condensed-matter
systems.\cite{hasan10,kane13} 
In a pioneering work in 1988, Haldane showed that a two-dimensional
graphene-like lattice model  with broken time-reversal symmetry
can exhibit an integer quantum Hall effect (IQHE) without an external
magnetic field.\cite{haldane88} 
Later, in 2005, Kane and Mele generalized this model to restore
time-reversal symmetry with the help of the natural spin degree of 
freedom in graphene with spin-orbit coupling.\cite{kane05, kane05-1}
The lowest-energy spin bands in this model carry non-zero but opposite 
Chern numbers that result in the quantum spin Hall effect (QSHE),
which manifests itself in a quantized conductance associated with
the transverse spin current. Whereas the intrinsic spin-orbit coupling is
too small in graphene to reveal the effect, the QSHE was later 
predicted\cite{bhz06} to occur and measured\cite{konig} in HgTe/CdTe
quantum wells.

The presence of an IQHE in a band with a non-zero Chern number
indicates a certain similarity between the band and the Landau level
of the two-dimensional electron gas (2DEG) in a strong magnetic
field. However, in contrast to 
the latter the energy bands obtained in tight-binding models 
have usually a non-negligible dispersion. In order to investigate in
further details the relation between Landau levels and energy bands 
with non-zero Chern numbers, special effort has recently been invested
into the engineering of flat bands in specially designed tight-binding 
models.\cite{tang11,sun11,neupert11} If these bands are partially
filled and if electron-electron interactions are taken into account,
one would then expect correlation effects similar to the fractional
quantum Hall effect (FQHE). The effect, also called fractional Chern
insulator,  was later corroborated within numerical
studies\cite{sheng11,regnault11}   
(for recent reviews, see \onlinecite{sondhi13,liu13,neupert13}).

The analogy between flat bands with non-zero Chern number and Landau
levels can be pushed further when the internal spin degree of freedom is taken
into account. Indeed, when there are as many electrons as flux quanta
threading the 2DEG, the spins are spontaneously aligned  
and form a ferromagnetic state (quantum Hall ferromagnet) in order to
minimize the electron-electron repulsion.\cite{ezawa} This situation 
corresponds to half-filling in a lattice model if only the two lowest
(spin) bands are taken into account. However, in contrast to Landau  
levels, where time-reversal symmetry is broken by the external
magnetic field and the Landau levels occur merely in two spin copies,
the situation is more involved in lattice models. One needs to
distinguish two generic situations. In the first one, time-reversal
symmetry is preserved such that spin and orbital degrees of
freedom are coupled. In this case, possible spin excitations are
described in the framework of rather unusual commutation relations for
the spin-density operators\cite{goerbig12} and the resulting
ferromagnetic state is expected to respect the underlying Z$_2$
symmetry of topological insulators.\cite{kane05} This type of
ferromagnetism has recently been investigated within numerical
exact-diagonalization studies by Neupert \textit{et al.}, who find a
gapped Ising ferromagnetic ground state.\cite{neupert12} The second
situation arises when time-reversal symmetry is broken on the level of
the lattice model, and where the Chern bands occur in two spin copies
without any spin-orbit structure.

In the present paper, we investigate ferromagnetism in both
situations, within a specially adapted tight-binding model on a square
lattice with on-site Hubbard repulsion that is a generalization of the
model originally presented in Ref.~\onlinecite{neupert11}. 
The tight-binding model, in the absence of interactions, bears a
staggered $\pi$-flux phase for each spin component, and time-reversal 
symmetry determines whether the two spin species experience either the same
or an opposite flux per plaquette. If time-reversal symmetry is broken, one is
confronted with a correlated flat-band Chern insulator, whereas one finds a 
correlated flat-band Z$_2$ topological insulator in the case of preserved
time-reversal symmetry. For both situations, we
investigate the ferromagnetic state at quarter-filling of the lattice that corresponds to
half-filling of the two lowest energy bands. In order to investigate its stability
and collective spin-wave excitations, we construct a nonpertubartive 
bosonization scheme similar to the one proposed in Ref. \onlinecite{doretto05}
to describe the 2DEG at filling factor $\nu = 1$.
Such a formalism was also applied to study   
quantum Hall ferromagnetic phases realized in graphene at filling factors
$\nu=0$ and $\nu=\pm1$\cite{doretto07} and to describe the
Bose-Einstein condensate  of magnetic excitons realized in a bilayer
quantum Hall system at total filling factor $\nu_T = 1$.\cite{doretto06,doretto12}

\begin{figure*}[t]
\centerline{\includegraphics[width=15.5cm]{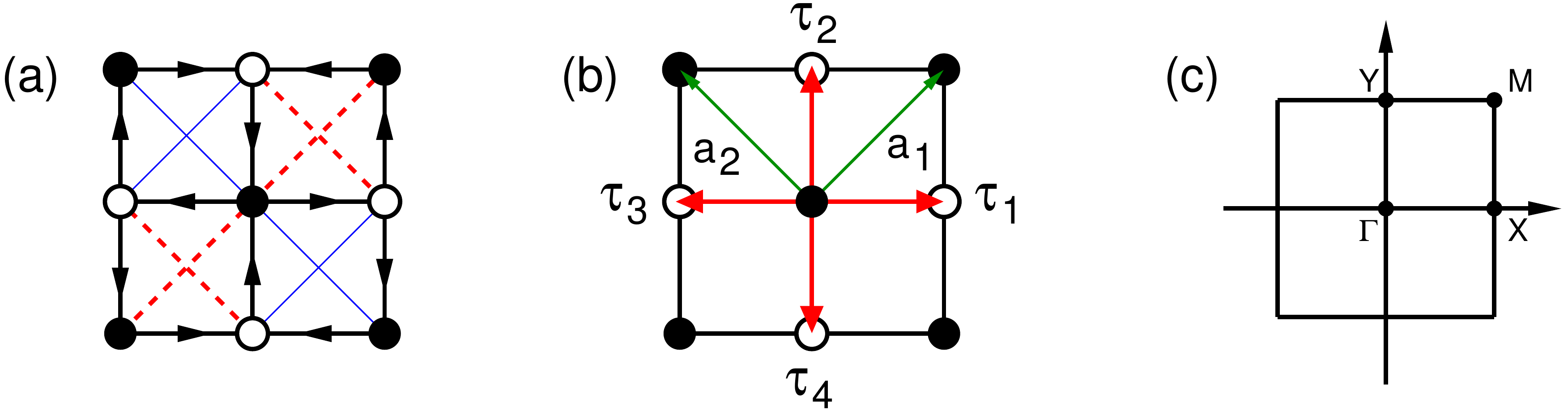}} 
\caption{(Color online)  
         (a) Schematic representation of the hopping term
         \eqref{ham-0}. The nearest-neighbor hopping energies (solid
         black lines) are equal
         to $t_1\exp\left( i\pi/4\right)$ (spin up electrons) in the
         direction of the arrows and the next-nearest-neighbor hopping
         energies are equal to $+t_2$ (dashed red lines) and 
         $-t_2$ (thin blue lines). Black and open circles 
         indicate sites of the $A$ and $B$ sublattices, respectively.  
         (b) The nearest-neighbor vectors $\taub_i$
         [read arrows, Eq.~\eqref{tauvectors}] and  
         the primitive vectors ${\mathbf a}_1  = a\hat{x}$ and 
         ${\mathbf a}_2  = a\hat{y}$ 
         [green arrows, Eq.~\eqref{deltavectors}].
         The next-nearest neighbor distance $a$ is set to one.
         (c) Brillouin zone. Here ${\bf X} = (\pi,0)$, ${\bf Y} = (0,\pi)$, and 
         ${\bf M} = (\pi,\pi)$.
}
\label{fig:lattice}
\end{figure*}

%%%%%%%%%%%%%%%%%%%%%%%%%%%%%%%%%%%%%%%%%%%%%%%%%%%%%%%%%%%%%%%
\subsection{Overview of the results}

We show that the bosonization scheme,\cite{doretto05} originally
developed for the 2DEG at filling factor $\nu=1$, can be
generalized for lattice models that describe flat-band Chern and Z$_2$
topological insulators.  

For both {\it correlated} Chern and Z$_2$ topological insulators described
above, we map the interacting fermion model to an effective {\it
  interacting} boson model.  We consider the effective boson model
in the harmonic approximation and show that the ground state is indeed
given by a spin polarized (ferromagnetic) state. 
Our main results are in fact the analytical calculation of the spin-wave
excitation spectra of both ferromagnets (Figs.~\ref{fig:disp-u} and
\ref{fig:disp-u-sqhe}). 
The spin-wave dispersion relation corresponds
to the energy of the bosons at the harmonic approximation.   
Due to the bipartite nature of the underlying square lattice, we identify two
types of collective spin-wave excitations.
We find that the correlated flat-band Chern insulator has one gapless
and one gapped spin-wave excitation branches (Fig.~\ref{fig:disp-u}) 
while the correlated flat-band topological insulator, two gapped ones
(Fig.~\ref{fig:disp-u-sqhe}). 
For the latter, the excitation gap we obtain  
at zero-wave vector coincides with the result numerically calculated by 
Neupert {\it et al.}\cite{neupert12}

%%%%%%%%%%%%%%%%%%%%%%%%%%%%%%%%%%%%%%%%%%%%%%%%%%%%%
\subsection{Outline}

Our paper is organized as follows. Section~\ref{sec:TBmodel}
introduces the basic tight-binding model (the spinfull square lattice
$\pi$-flux model), which is discussed in view of the role played by
time-reversal symmetry. We discuss the flat-band 
ferromagnetic phase of a correlated Chern insulator with broken
time-reversal symmetry in Sec.~\ref{sec:hall}, whereas the more
involved case of a model with underlying time-reversal symmetry
(correlated topological insulator) is presented in Sec.~\ref{sec:shall}.  
For both cases, the particularities of the associated lattice models
are first discussed in Secs.~\ref{sec:model0}  and \ref{sec:model02} before
we present the details of the bosonization schemes in Sec.~\ref{sec:boso} 
and Sec.~\ref{sec:boso-sqhe}, respectively.
The bosonization formalism is then applied in
Secs.~\ref{sec:boso-model} and \ref{sec:boso-model02}  
to study the flat-band ferromagnetic phases obtained in the presence
of a strong on-site Hubbard repulsion term in the respective models.  
We comment on possible extensions of the bosonization scheme and the
effects of the boson-boson (spin-wave--spin-wave) coupling in Sec.~\ref{sec:discussion}
and, in Sec.~\ref{sec:summary}, we provide a brief summary of our
findings. 
Technical details of the two bosonization schemes, as well as a more
detailed analysis of time-reversal symmetry, are delegated to the
(three) Appendices.

%%%%%%%%%%%%%%%%%%%%%%%%%%%%%%%%%%%%%%%%%%%%%%%%%%%%%%%%%%%%%%%%%%%%%%%%%%%%%%%%%%%%%
\section{Tight-binding models with flat topological bands}
\label{sec:TBmodel}

Before a detailed analysis of the different ferromagnetic states in
flat-band Chern insulators with broken time-reversal symmetry  
and Z$_2$ topological insulators, we present here a common
tight-binding model that provides the different flat bands.

Let us consider $N_e = N$ free spin-$1/2$ electrons hopping on a
bipartite square lattice where both sublattices $A$ and $B$ have 
each $N_A = N_B = N$ sites. The Hamiltonian of the system is given by
the tight-binding model
\begin{eqnarray}
 \mathcal{H}_0 &=& \sum_{ i \in A, \, n, \, \sigma}
                 \left( t_{i,i+n,\sigma} \,c^\dagger_{i\, A\, \sigma}c_{i+n\, B\, \sigma} + {\rm H.c.} \right)
\nonumber \\
              &+& \sum_{i,\,\delta,\,a,\,\sigma}
                 \left( \lambda_{i,i+\delta}\,c^\dagger_{i\, a\, \sigma}c_{i+\delta\, a\, \sigma} + {\rm H.c.} \right).
\label{ham-0}
\end{eqnarray}
Here $c^\dagger_{i\, a\, \sigma}$ ($c_{i\, a\, \sigma}$) creates (destroys) a
spin $\sigma = \uparrow,\downarrow$ electron on site $i$ of the
sublattice $a= A, B$. 
The {\it spin-dependent} nearest-neighbor $t_{i,j,\sigma}$ and 
next-nearest-neighbor  $\lambda_{i,j}$ hopping energies are
respectively given by [Fig.~\ref{fig:lattice}(a)]
\begin{equation}
   t_{i,i+n,\sigma} = 
     t_1\exp\left[ i(-1)^n\gamma(\sigma)\pi/4\right], \;\;\;\;  i \in A,
\label{hopping1}
\end{equation}
\begin{equation}
   \lambda_{i,i+\delta} = \left\{ \begin{tabular}{ll}
                                $-(-1)^\delta t_2$, & $\;\; i\in A$, \\ 
                                $+(-1)^\delta t_2$, & $\;\; i\in B$,  
                                \end{tabular}
                          \right.
\label{hopping2}
\end{equation}
where $t_1$ and $t_2$ are positive real quantities. The indices 
$n = 1,2,3,4$ correspond to the nearest-neighbor vectors [Fig.~\ref{fig:lattice}(b)]
\begin{eqnarray}
 \taub_1 &=& - \taub_3 = \frac{1}{2}\left(  {\mathbf a}_1 - {\mathbf a}_2 \right)   
                                      = \frac{a}{2}(\hat{x} - \hat{y}),  \nonumber \\
&& \label{tauvectors} \\
 \taub_2 &=& - \taub_4 = \frac{1}{2}\left(  {\mathbf a}_1 + {\mathbf a}_2 \right) 
                                      = \frac{a}{2}(\hat{x} + \hat{y}), \nonumber 
\end{eqnarray}
and $\delta$ is either $1$ or $2$ corresponding to the next-nearest-neighbor vectors
\begin{equation}
 \deltab_1 = {\mathbf a}_1  = a\hat{x}, \;\;\;\;\;\;\;
 \deltab_2 = {\mathbf a}_2  = a\hat{y}.
\label{deltavectors}
\end{equation}
In the remainder of this paper, we set the next-nearest-neighbor distance $a=1$.
The spin-dependent phases $\gamma(\sigma)$ reflect that time-reversal
symmetry is preserved for 
$\gamma(\uparrow) = -\gamma(\downarrow) = 1$,
whereas it is broken if  
$\gamma(\uparrow) = \gamma(\downarrow) = 1$
(see Appendix~\ref{ap:symmetries} for details).  
The tight-binding model \eqref{ham-0} is a generalization for the case
of electrons with spin of the $\pi$-flux model discussed  in Ref.~\onlinecite{neupert11}. 
Since the nearest-neighbor hopping energy \eqref{hopping1} is
complex, each electron acquires a phase $\pi$ as it hops around a
plaquette in the direction of the arrows indicated in
Fig.~\ref{fig:lattice}(a). Therefore, $\mathcal{H}_0$ describes
noninteracting electrons hopping in a square lattice in the presence
of a fictitious staggered $\pm\pi$ flux pattern.\cite{lee06} 
For the time-reversal-symmetric model [$\gamma(\uparrow) = -\gamma(\downarrow) = 1$] 
the flux experienced by electrons of opposite spin is opposite, whereas it is the same
for $\gamma(\uparrow) = \gamma(\downarrow) = 1$, i.e., in the case of broken time-reversal
symmetry.

After Fourier transformation, i.e., introducing
\begin{equation}
 c^\dagger_{i\, a\,\sigma} = \frac{1}{N^{1/2}_a}\sum_{\bk \in {\rm BZ}} 
          \exp(-i\bk\cdot\bR_i)c^\dagger_{\bk\, a\,\sigma}
\label{fourier}
\end{equation}
with the momentum sum running over the BZ associated
with the underlying Bravais lattice [see Fig.~\ref{fig:lattice}(c)], 
it is possible to show that the hopping term \eqref{ham-0} assumes the
form 
\begin{equation}
\mathcal{H}_0 = \sum_{\bk \in BZ}\Psi^\dagger_\bk\mathcal{H}_\bk\Psi_\bk,
\label{ham-0-k}
\end{equation}
where
\begin{equation}
\mathcal{H}_k = \left( 
  \begin{array}{cc} h^\uparrow_\bk & 0 \\
                              0 & h^\downarrow_\bk 
  \end{array} \right)
\label{hk} 
\end{equation}
is a $4\times 4$ matrix and
\begin{equation}
 \Psi^\dagger_\bk = 
    \left( c^\dagger_{\bk\,A\,\uparrow} \;\; 
             c^\dagger_{\bk\,B\,\uparrow} \;\;
             c^\dagger_{\bk\,A\,\downarrow} \;\; 
             c^\dagger_{\bk\,B\,\downarrow}
     \right)
\end{equation}
is a four-component spinor. Furthermore,
\begin{equation}
  h^\uparrow_\bk = B_{0,\bk}\tau_0 + \bB_\bk\cdot\hat{\tau}
\label{hkup}
\end{equation}
is a $2\times 2$ matrix  where $\tau_0 = I_{2\times 2}$ is the
identity matrix and $\hat{\tau} = (\tau_1,\tau_2,\tau_3)$ is 
a vector whose components are the Pauli matrices.
Finally, $B_{0,\bk}$ and the components of
the vector $\bB_\bk = (B_{1,\bk},B_{2,\bk},B_{3,\bk})$ are given by
the functions  
\begin{eqnarray}
 B_{0,\bk} &=& 0,
\nonumber \\
 B_{1,\bk}  &=& 2\sqrt{2}t_1\cos\frac{k_x}{2}\cos\frac{k_y}{2}, 
\nonumber \\
 B_{2,\bk}  &=& 2\sqrt{2}t_1\gamma(\sigma)\sin\frac{k_x}{2}\sin\frac{k_y}{2},
\nonumber \\
 B_{3,\bk} &=& 2t_2\left(\cos k_x - \cos k_y \right).
\label{b-coef}
\end{eqnarray}      
Again, the factor $\gamma(\sigma)$ indicates whether time-reversal
symmetry is broken or not.  Whereas the time-reversal-symmetric model
[$\gamma(\uparrow) = -\gamma(\downarrow) = 1$] has the usual property
$h_{\bk}^{\uparrow}=(h_{-\bk}^{\downarrow})^*$, we find that the two
components of the Hamiltonian  \eqref{hk} are identical for all wave vectors, 
$h^\downarrow_\bk = h^\uparrow_\bk$, in the case of 
$\gamma(\uparrow) = \gamma(\downarrow) = 1$.

%%%%%%%%%%%%%%%%%%%%%%%%%%%%%%%%%%%%%%%%%%%%%%%%%%%%%%%%%%%%%%%%%%%%%%%%%%%%%%%%%%%%%
\subsection{Symmetries of the $\pi$-flux model: spin rotation}
\label{sec:symmetries}

The above discussion and the role of time-reversal symmetry allow us
to investigate certain properties of the supposed flat-band
ferromagnetic states from a pure symmetry point of view.  
In this section, we discuss the behavior of the noninteracting
fermion model \eqref{ham-0} under spin rotation. Some further considerations
about the behavior of \eqref{ham-0} under time-reversal are present in
Appendix~\ref{ap:symmetries}.  

The Hamiltonian \eqref{ham-0-k} can indeed be written as 
\begin{eqnarray}
 \mathcal{H}_0 &=& \sum_{\bk,\sigma} 
   \sum_{a,b} E^{ab}_\sigma(\bk)c^\dagger_{\bk\,a\,\sigma}c_{\bk\,b\,\sigma}. 
\label{aux-sym1}
\end{eqnarray}
Comparing Eq.~\eqref{aux-sym1} with Eqs.~\eqref{hk} and \eqref{hkup},
we see that, for the Chern insulator with broken time-reversal symmetry 
($h^\uparrow_\bk = h^\downarrow_\bk$),  
\begin{equation}
  E^{aa}_\uparrow(\bk) = E^{aa}_\downarrow(\bk)
  \;\;\;\; {\rm and} \;\;\;\;
   E^{ab}_\uparrow(\bk) =  E^{ab}_\downarrow(\bk)
\label{ham-chern}
\end{equation}
with $a\not= b$, while for the topological insulator with
time-reversal symmetry
($h^\downarrow_\bk = h^{\uparrow\,*}_{-\bk}$), we have
\begin{equation}
  E^{aa}_\uparrow(\bk) = E^{aa}_\downarrow(\bk)
  \;\;\;\;{\rm and}\;\;\;\;
   E^{ab}_\uparrow(\bk) \not=  E^{ab}_\downarrow(\bk)
\label{ham-top}
\end{equation}
with $a\not=b$.

The total spin operator reads
\begin{equation}
\bS = \sum_{i,a}\bS_{i\,a} = \frac{1}{2}\sum_{a,\bp}
 c^\dagger_{\bp\,a\,\sigma}\hat{\sigma}_{\sigma,\sigma'}c_{\bp\,a\,\sigma'}
\label{eq:TotSpin}
\end{equation}
where $\bS_{i\,a}$ is the spin operator at site $i$ of the sublattice
$a$ and  $\hat{\sigma}=(\sigma_1,\sigma_2,\sigma_3)$ is a 
vector of Pauli matrices associated with the physical spin 
[see Eq.~\eqref{spin-op} below]. 
It is possible to show that the following commutation relations hold:
\begin{eqnarray}\label{eq:comms}
 [\mathcal{H}_0,S^z] &=& 0
\nonumber \\
&& \nonumber \\
\, [\mathcal{H}_0, S^+] &=& \sum_\bk 
  \sum_a \left[E^{aa}_\uparrow(\bk)  -  E^{aa}_\downarrow(\bk) \right] 
         c^\dagger_{\bk,\,a\,\uparrow}c_{\bk,\,a\,\downarrow}
\nonumber \\
&& \nonumber \\
&& + \left[  E^{AB}_\uparrow(\bk) -  E^{AB}_\downarrow(\bk)\right] 
         c^\dagger_{\bk,\,A\,\uparrow}c_{\bk,\,B\,\downarrow}
\nonumber \\
&& \nonumber \\
&&   + \left[  E^{BA}_\uparrow(\bk) -  E^{BA}_\downarrow(\bk) \right] 
         c^\dagger_{\bk,\,B\,\uparrow}c_{\bk,\,A\,\downarrow},
\nonumber \\
&& \nonumber \\
\, [\mathcal{H}_0, S^-] &=& [S^+, \mathcal{H}_0]^*. 
\end{eqnarray}

Since the Chern insulator is characterized by
the Hamiltonian \eqref{hk} with two identical components for the two
spin orientations [see Eq.~\eqref{ham-chern}], one immediately realizes 
that all components of the total spin operator \eqref{eq:TotSpin}
commute with the Hamiltonian \eqref{hk}, i.e., the Hamiltonian has
SU(2) spin rotation symmetry.
In the case of a ferromagnetic ground state that breaks spin rotation
symmetry, all equivalent states can thus be obtained by global
rotations generated by the total spin operator \eqref{eq:TotSpin}, and
one would therefore expect the presence of a Goldstone mode in the
spin-wave excitation spectrum. We show explicitly in the following
section the existence of such a mode. 

Concerning a topological insulator described by the Hamiltonian
\eqref{hk} with coefficients obeying Eq.~\eqref{ham-top}, one can
easily see that the SU(2) spin rotation symmetry is now explicitly broken to
U(1), i.e., the Hamiltonian is invariant under spin rotations around
the $z$-axis.
Therefore, the presence of a Goldstone mode depends
on the ground-state spin polarization: whereas one would expect a
superfluid-type mode for an {\it easy-plane} 
ferromagnetic state, where the polarization is oriented in the
$xy$-plane, an {\it easy-axis} ferromagnet with a polarization along the
$z$-direction would not display a Goldstone mode since the ground
state preserves the U(1) spin rotation symmetry of the Hamiltonian. In
the following, we show that the latter is the case for a topological
insulator and that all collective excitations are indeed gapped.

%%%%%%%%%%%%%%%%%%%%%%%%%%%%%%%%%%%%%%%%%%%%%%%%%%%%%%%%%%%%%%%%%%%%%%%%%%%%%%%%%%%%%
\section{Ferromagnetism in a flat-band Chern insulator}
\label{sec:hall}

In this section, we discuss the flat-band ferromagnetic phase
of a topological Hubbard model that explicitly breaks time-reversal
symmetry. We start by discussing the free electron term of the
model.

%%%%%%%%%%%%%%%%%%%%%%%%%%%%%%%%%%%%%%%%%%%%%%%%%%%%%%%%%%%%%%%%%%%%%%%%%%%%%%%%%%%%%
\subsection{Square lattice $\pi$-flux model with broken time-reversal symmetry}
\label{sec:model0}

The noninteracting Hamiltonian \eqref{ham-0-k} with
$\gamma(\uparrow)=\gamma(\downarrow)=1$ can be diagonalized
with the aid of the Bogoliubov transformation 
\begin{eqnarray}
 c^\dagger_{\bk\,A\,\sigma} &=& u_\bk d^\dagger_{\bk\, \sigma} 
                     + v^*_\bk c^\dagger_{\bk\,  \sigma},
\nonumber \\
 c^\dagger_{\bk\,B\,\sigma} &=& v_\bk d^\dagger_{\bk\,\sigma} 
                     - u^*_\bk c_{\bk\,\sigma},
\label{bogo-qhe}
\end{eqnarray}
where the Bogoliubov coefficients $u_\bk$ and $v_\bk$ are given by
Eq.~\eqref{coef-bogo}. 
After diagonalization, the Hamiltonian \eqref{ham-0-k} assumes the
form [see Eqs.~\eqref{coef-bogo}-\eqref{coef-bogo2} for details]
\begin{equation}
\mathcal{H}_0 = \sum_{\bk \in BZ}\Phi^\dagger_\bk\mathcal{H'}_\bk\Phi_\bk,
\label{diag-ham-0}
\end{equation}
where the $4\times 4$ matrix $\mathcal{H'}_\bk$ reads
\begin{equation}
\mathcal{H'}_\bk = \left( 
  \begin{array}{cc} h'_\bk & 0 \\
                              0 & h'_\bk 
  \end{array} \right),
\end{equation}
with 
\begin{equation}
 h'_\bk = \left( 
  \begin{array}{cc}  \omega_{d,\bk} & 0 \\
                              0 & \omega_{c,\bk}
  \end{array} \right)
\label{h-matrix}
\end{equation}
being a $2\times 2$ diagonal matrix whose elements are the upper band
$d$ ($+$ sign) and the lower one $c$ ($-$ sign),   
\begin{equation}
  \omega_{d/c,\bk} = B_{0,\bk} \pm |\bB_\bk|,
\label{omega-cd}
\end{equation}
and the new four component spinor $\Phi^\dagger_\bk$ is given by 
\begin{equation}
 \Phi^\dagger_\bk = 
    \left( d^\dagger_{\bk\,\uparrow}  \;\; 
            c^\dagger_{\bk\,\uparrow} \;\;
            d^\dagger_{\bk\,\downarrow}\;\;
            c^\dagger_{\bk\,\downarrow}
   \right).
\end{equation}

\begin{figure}[t]
\centerline{\includegraphics[width=9.7cm]{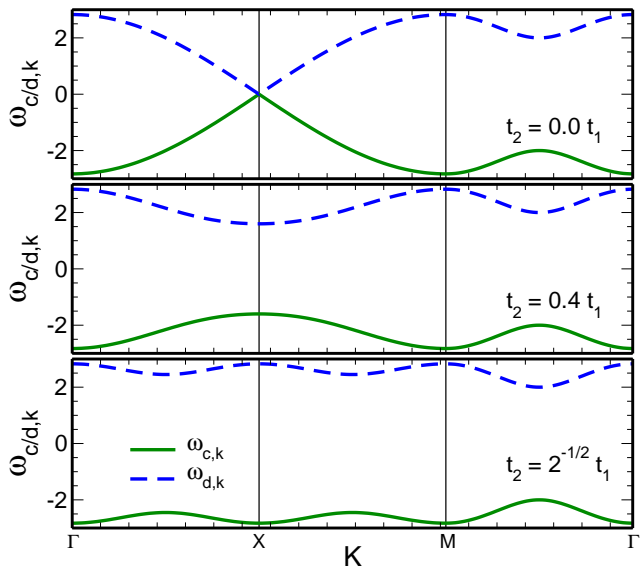}} 
\caption{(Color online) Band structure of the noninteracting hopping
  term $\mathcal{H}_0$ [Eq.~\eqref{omega-cd} in units of $t_1$] 
  along paths in the Brillouin zone
  [Fig.~\ref{fig:lattice}(c)] for different values of the 
  next-nearest-neighbor hopping energy amplitude $t_2$. Solid green
  and dashed blue lines correspond respectively to the $c$ and $d$ bands.   
}
\label{fig:bands}
\end{figure}

\begin{figure}[t]
\centerline{\includegraphics[width=8.5cm]{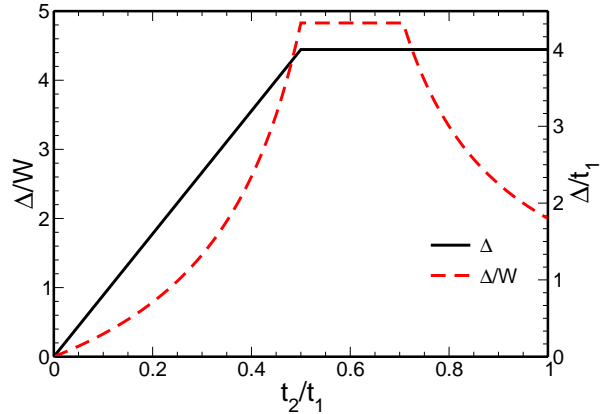}} 
\caption{(Color online)  Flatness ratio $\Delta/W$ of the $c$-band 
              and band gap $\Delta$ as a function of the 
              next-nearest-neighbor hopping energy amplitude
             $t_2$ for the noninteracting hopping term $\mathcal{H}_0$.   
}
\label{fig:gap}
\end{figure}

The band structure of $\mathcal{H}_0$ is made out of four bands: 
\[ 
    d\uparrow, \,\,  d\downarrow,\,\,
    c\uparrow, \,\,  c\downarrow, 
\]
with the $c$ and $d$ bands doubly degenerate in the spin degree of
freedom as expected from the discussion of the previous
section. Figure~\ref{fig:bands} 
shows the energy of the $c$ and $d$ bands [Eq.~\eqref{omega-cd}] for
different values of the ratio $t_2/t_1$.  Note that the spectrum is
gapless for $t_2 = 0$ and, as $t_2$ increases, it acquires a finite energy gap  
$\Delta = {\rm min}(\omega_{d,\bk}) - {\rm max}(\omega_{c,\bk})$.
Indeed, as shown in Fig.~\ref{fig:gap}, $\Delta$ linearly increases
with $t_2$ for $t_2 < 0.5\,t_1$ and then it saturates at $\Delta = 4\,t_1$ 
for larger values of $t_2$. The widths 
$W_{c/d} = {\rm max}(\omega_{c/d,\bk}) - {\rm min}(\omega_{c/d,\bk})$
of the $c$ and $d$ bands also change with $t_2$. In particular,
as illustrated in Fig.~\ref{fig:gap}, the flatness ratio 
$f_c = \Delta/W_c$ of the $c$-band\cite{sondhi13} is maximum 
($f_c = 4.83$) for $0.5 \le t_2/t_1 \le 1/\sqrt{2}$.  Such a range
includes the configuration $t_2 = t_1/\sqrt{2}$ discussed by
Neupert {\it et al}.\cite{neupert11}     

An interesting aspect of the tight-binding model $\mathcal{H}_0$ is
that its energy bands are topologically nontrivial. Indeed, 
it is possible to show that the Chern numbers of the $c$ and $d$ bands 
can be written in terms of the coefficients $B_{i,\bk}$ [Eq.~\eqref{b-coef}] 
and that they are finite, i.e.,\cite{hasan10,kane13,neupert11,assaad13}   
\begin{equation}
  C^{d/c}_\sigma =  \mp\frac{1}{4\pi}\int_{BZ}d^2k \;\;
   \hat{B}_\bk\cdot(\partial_{k_x}\hat{B}_\bk\times\partial_{k_y}\hat{B}_\bk)
      = \mp 1,   
\label{chern-number2} 
\end{equation}
with $\hat{\bB}_\bk \equiv \bB_\bk / |\bB_\bk|$, regardless the value
of $t_2 > 0$. Therefore, the square lattice $\pi$-flux model \eqref{ham-0}
with broken time-reversal symmetry is an example of a Chern
insulator. As discussed in the Introduction, the system should exhibit
an IQHE when a certain number of energy bands are completely filled.

In the following, we focus on the nearly flat-band limit of
$\mathcal{H}_0$, in particular, we consider $t_2 = t_1/\sqrt{2}$.
Note that since each of the $c\,\sigma$ and $d\,\sigma$ bands have 
$N_A = N_B = N $ available states and $N_e = N$, the $d$ band is empty while
the $c$ one is half-filled, i.e., we have $1/4$-filling including all four bands.
In the next section, we introduce a bosonization scheme to describe
such a flat-band Chern insulator. 

%%%%%%%%%%%%%%%%%%%%%%%%%%%%%%%%%%%%%%%%%%%%%%%%%%%%%%%%%%%%%%%%%%%%%%%%%%%%%%%%%%%%%
\subsection{Bosonization scheme for a flat-band Chern insulator}
\label{sec:boso}

In order to study the flat-band ferromagnetic phase of a correlated 
Chern insulator, we introduce a bosonization scheme similar to that
proposed in Ref.~\onlinecite{doretto05} for the 2DEG at $\nu=1$.
Following the lines of the bosonization method for one-dimensional
fermion systems,\cite{giamarchi} the idea of the formalism\cite{doretto05} 
is to define boson operators in terms of the lowest energy neutral 
excitations of the original fermionic system and then map the 
interacting electronic model to an effective bosonic one.      
In the following, we outline the bosonization scheme for flat-band Chern
insulators. The (two) approximations involved and the differences
between the scheme on a lattice and the one developed for the 
2DEG at $\nu=1$ are discussed. 
Further details can be found in Appendix~\ref{ap:boso}.

Let us consider the tight-binding model \eqref{ham-0}  at
$1/4$-filling. As mentioned in the previous section, in this case the
highest energy $d$ bands are completely empty while the lowest energy $c$
ones are half-filled, see Fig.~\ref{fig:bands}. In particular, let us assume that 
the $c\uparrow$ band is completely filled while the $c\downarrow$ one is  
completely empty, i.e., we consider that the ground state of the free
electron model \eqref{ham-0} is completely spin polarized,   
\begin{equation}
 |{\rm FM}\rangle = \prod_{\bk \in {\rm BZ}}c^\dagger_{\bk\,\uparrow}|0\rangle.
\label{fm} 
\end{equation}
The state $|{\rm FM}\rangle $ is the reference state of the
bosonization method and belongs to the ground-state manifold of ferromagnetic
states that need to be considered when electronic interactions are taken into account.
Since the lowest energy $c$ bands are separated from the highest
energy $d$ ones by an energy gap and the former bands are
partially filled, the lowest-energy neutral excitations are given by
particle-hole pairs (spin flips) within the $c$ bands. 
Therefore, in the following, we neglect the $d$ bands,
i.e., we restrict the Hilbert space to the subspace spanned by the
lowest energy $c$ bands.

The spin operator at site $i$ of the sublattice $a$ is defined as 
\begin{equation}
\bS_{i\,a} = \frac{1}{2}c^\dagger_{i\, a\,\sigma}\hat{\sigma}_{\sigma\,\sigma'}
                   c_{i\, a\,\sigma'}
\label{spin-op}
\end{equation}
in terms of the Pauli matrices defined above. 
The Fourier transform of the components of $\bS_{i\,a}$ read 
\begin{equation}
 S^\lambda_{i\,a} = \frac{1}{N_a}\sum_\bq\exp(i\bq\cdot\bR_i)S^\lambda_{\bq,a}
\label{fourier-spin}
\end{equation}
with $\lambda = x,y,z$. In particular, using Eq.~\eqref{fourier}, it is
possible to show that  
\begin{eqnarray}
 S^+_{\bq,a} &=& \sum_\bp c^\dagger_{\bp-\bq\, a\,\uparrow}c_{\bp\, a\,\downarrow},
\nonumber \\
 S^-_{\bq,a} &=& \sum_\bp c^\dagger_{\bp-\bq\, a\,\downarrow}c_{\bp\, a\,\uparrow},
\end{eqnarray}
where $S^\pm_{\bq,a} = S^x_{\bq,a}  \pm iS^y_{\bq,a}$. With the help of
the Bogoliubov transformation \eqref{bogo-qhe}, one can express the
spin operators $S^\pm_{\bq,a}$ in terms of the $c$ and $d$ fermion
operators. For instance,   
\begin{eqnarray}
 S^+_{\bq,A} &=& \sum_\bp \left(
            u_{\bp-\bq}u^*_\bp d^\dagger_{\bp-\bq\,\uparrow}d_{\bp\,\downarrow}
          + v^*_{\bp-\bq}v_\bp c^\dagger_{\bp-\bq\,\uparrow}c_{\bp\,\downarrow}
         \right.      
\nonumber \\ 
&& \nonumber \\
&+& 
       \left. 
        u_{\bp-\bq}v_\bp d^\dagger_{\bp-\bq\,\uparrow}c_{\bp\,\downarrow}
        + v^*_{\bp-\bq}u^*_\bp c^\dagger_{\bp-\bq\,\uparrow}d_{\bp\,\downarrow}
            \right), \;
\label{s+a}
\end{eqnarray}
where $u_\bp$ and $v_\bp$ are the Bogoliubov coefficients \eqref{coef-bogo}.
Note that by projecting $S^+_{\bq,A}$ into the $c$ bands, only the second
term on the R.H.S. of Eq.~\eqref{s+a} survives.
Indeed, it is easy to see that the expressions of the {\it projected} spin operators
$\bar{S}^\pm_{\bq,a}$ read 
\begin{eqnarray}
 \bar{S}^+_{\bq,a} &=& \sum_\bp
           G_a(\bp,\bq)c^\dagger_{\bp-\bq\,\uparrow}c_{\bp\,\downarrow},
\nonumber \\
 \bar{S}^-_{\bq,a} &=& \sum_\bp
          G_a(\bp,\bq)c^\dagger_{\bp-\bq\,\downarrow}c_{\bp\,\uparrow},
\label{aux-spin-op}
\end{eqnarray}
where 
\begin{equation}
  G_A(\bp,\bq) = v^*_{\bp-\bq}v_\bp 
 \;\;\;\; {\rm and} \;\;\;\;
  G_B(\bp,\bq) = u^*_{\bp-\bq}u_\bp. 
\label{gab-functions}
\end{equation}
As discussed below, it is interesting to consider the
following linear combination of the spin operators
$\bar{S}^\lambda_{\bq,a}$:    
\begin{equation}
 \bar{S}^\lambda_{\bq,\alpha}  =  \bar{S}^\lambda_{\bq,A} 
                                          + (-1)^\alpha\bar{S}^\lambda_{\bq,B} 
\label{spin-alpha-op}
\end{equation}
with $\lambda = x,y,z$ and $\alpha = 0,1$. We have, for instance,  
\begin{eqnarray}
 \bar{S}^+_{\bq,\alpha} &=& \sum_\bp 
          g_\alpha(\bp,\bq)c^\dagger_{\bp-\bq\,\uparrow}c_{\bp\,\downarrow},
\nonumber \\
 \bar{S}^-_{\bq,\alpha} &=& \sum_\bp
          g_\alpha(\bp,\bq)c^\dagger_{\bp-\bq\,\downarrow}c_{,\bp\,\uparrow},
\label{spin-alpha-pm-op}
\end{eqnarray}
where the $g_\alpha(\bp,\bq) $ functions are defined as 
\begin{equation}
  g_\alpha(\bp,\bq) = v^*_{\bp-\bq}v_\bp + (-1)^\alpha u^*_{\bp-\bq}u_\bp.
\label{g-func-hall}
\end{equation}

Similar considerations hold for the density operator of electrons with
spin $\sigma$ at site $i$, which is define as
\begin{equation}
  \hat{\rho}_{i\, a\, \sigma} = c^\dagger_{i\, a\, \sigma}c_{i\, a\, \sigma}.
\label{density-op}
\end{equation}
Its Fourier transform is given by 
\begin{equation}
  \hat{\rho}_{i\, a\, \sigma} = \frac{1}{N}\sum_{\bq\in {\rm BZ}}
         \exp(i\bq\cdot\bR_i) \hat{\rho}_{a \sigma}(\bq)
\end{equation}
and, with the help of Eq.~\eqref{fourier}, it is easy to see that 
\begin{equation}
  \hat{\rho}_{a \sigma}(\bq) = \sum_\bp c^\dagger_{\bp-\bq\, a\,\sigma}c_{\bp\, a\,\sigma}.
\label{density-op-2}
\end{equation}
Finally, the projected electron density operator reads
\begin{equation}
 \bar{\rho}_{a\, \sigma}(\bq) = \sum_\bp 
    G_a(\bp,\bq)c^\dagger_{\bp-\bq\,\sigma}c_{\bp\,\sigma} 
\label{proj-dens-op} 
\end{equation}
with the $ G_a(\bp,\bq)$ functions given by Eq.~\eqref{gab-functions}.

Differently from the  Girvin-MacDonald-Platzman (GMP)
algebra\cite{gmp} for electrons within the lowest Landau level here,
for a flat-band Chern insulator, the algebra of the projected spin
[Eq.~\eqref{spin-alpha-op}] and electron density
[Eq.~\eqref{proj-dens-op}] operators is not closed. 
For instance, the commutator     
\begin{eqnarray}
[\bar{S}^+_{\bq,\alpha}, \bar{S}^-_{\bq',\beta}] &=& \sum_\bp  
            \left[ g_\alpha(\bp-\bq',\bq)g_\beta(\bp,\bq')
                c^\dagger_{\bp-\bq-\bq'\,\uparrow}c_{\bp\,\uparrow}
           \right.
\nonumber \\
&-&  \left.    g_\alpha(\bp,\bq)g_\beta(\bp-\bq,\bq')
             c^\dagger_{\bp-\bq-\bq'\,\downarrow}c_{\bp\,\downarrow}   \right]
\label{comut-ss} 
\end{eqnarray}
cannot be expressed in terms of the projected spin and electron
density operators [see, for instance,  Eq.~(37) from Ref.~\onlinecite{doretto05}].     
Similar considerations hold for the commutators
\[
 [ \bar{\rho}_{a\,\sigma}(\bk), \bar{S}^+_{\bq,\alpha}  ],
\;\;\;\;
 [ \bar{\rho}_{a\,\sigma}(\bk), \bar{S}^-_{\bq,\alpha}  ], 
\;\;\;\;
 [ \bar{\rho}_{a\,\sigma}(\bk), \bar{\rho}_{b\,\sigma'}(\bq) ],
\]
see Eqs.~\eqref{comut-s+rho}-\eqref{comut-rhorho}.
We refer the reader to
Refs.~\onlinecite{parameswaran12,goerbig12,roy14} for a discussion 
about the flat Berry {\it curvature} limit, where the GMP algebra
holds for flat-band Chern insulators. 

In order to define boson operators in terms of the
fermion operators $c$, we consider the neutral particle-hole pair
excitations above the ground state  $|{\rm FM}\rangle$ of the
free-electron model \eqref{diag-ham-0}. 
Such excitations, which correspond to spin--flips, are given by 
\begin{equation}
 |\Psi_\bq\rangle = \bar{S}^-_{\bq,\alpha}|{\rm FM}\rangle, 
 \;\;\;\;\;\;\;\;\;\;\;
 \alpha = 0,1,
\label{excitation-chern}
\end{equation}
where $\bar{S}^-_{\bq,\alpha}$ is the  linear combination \eqref{spin-alpha-op} 
of the spin operators $\bar{S}^-_{\bq,A}$ and $\bar{S}^-_{\bq,B}$.  
As shown below, the fermionic representation of the spin operators
$\bar{S}^\pm_{\bq,\alpha}$ allows us to define two sets of
independent boson operators.

The commutator \eqref{comut-ss} between the spin operators
$\bar{S}^+_{\bq,\alpha}$ and $\bar{S}^-_{\bq,\beta}$  differs from the usual canonical
commutation relation between creation and annihilation boson
operators. However, if the number of particle--hole pair excitations
is small, one can write 
\begin{eqnarray}
 c^\dagger_{\bp-\bq\,\uparrow}c_{\bp\,\uparrow} &\approx& 
  \langle {\rm FM} | c^\dagger_{\bp-\bq\,\uparrow}c_{\bp\,\uparrow} | {\rm FM} \rangle
       =  \delta_{\bq,0},
\nonumber \\
   c^\dagger_{\bp-\bq\,\downarrow}c_{\bp\,\downarrow} &\approx&
  \langle {\rm FM} | c^\dagger_{\bp-\bq\,\downarrow}c_{\bp\,\downarrow} | {\rm FM} \rangle
  =  0.
\label{assumption}
\end{eqnarray}
In this case, the commutator \eqref{comut-ss} acquires the form
\begin{equation}
  [\bar{S}^+_{\bq,\alpha}, \bar{S}^-_{\bq',\beta}] \approx  
                  \delta_{\bq,-\bq'}\sum_\bp g_\alpha(\bp -
                  \bq',-\bq')g_\beta(\bp,\bq'). 
\end{equation}
Moreover, it is possible to show that for the square lattice $\pi$-flux 
model \eqref{diag-ham-0} the sum over momentum in the
above equation is finite only if $\alpha=\beta$, 
see Eq.~\eqref{f-alpha-beta} for details, i.e., 
\begin{equation}
  [\bar{S}^+_{\bq,\alpha}, \bar{S}^-_{\bq',\beta}] \approx  
                  \delta_{\bq,-\bq'}\delta_{\alpha,\beta}
                  \sum_\bp g_\alpha(\bp - \bq',-\bq')g_\alpha(\bp,\bq').
\label{approx-comut} 
\end{equation}
Therefore, as long as the number of particle-hole pair excitations 
above the reference state $|{\rm FM}\rangle$ is small, the commutator
\eqref{comut-ss} is approximately equal to a canonical boson
commutation relation. In other words, in this limit, 
the lowest-energy particle-hole pair excitations can be approximately
treated as bosons.
We then {\it define} two sets of independent boson operators
\begin{eqnarray}
b_{\alpha,\bq} &=& \frac{\bar{S}^+_{-\bq,\alpha}}{F_{\alpha,\bq}}
  = \frac{1}{F_{\alpha,\bq}}
  \sum_\bp g_\alpha(\bp,-\bq)c^\dagger_{\bp+\bq\,\uparrow}c_{\bq\,\downarrow},
\nonumber\\
&& \nonumber \\
b^\dagger_{\alpha,\bq} &=& \frac{\bar{S}^-_{\bq,\alpha}}{F_{\alpha,\bq}}
  = \frac{1}{F_{\alpha,\bq}}
  \sum_\bp g_\alpha(\bp,\bq)c^\dagger_{\bp-\bq\,\downarrow}c_{\bq\,\uparrow},
\label{b-op}
\end{eqnarray}
with $\alpha = 0,1$, that obey the canonical commutation
relations
\begin{eqnarray}
   [b_{\alpha,\bk},b^\dagger_{\beta,\bq}] &=& \delta_{\alpha,\beta}\delta_{\bk,\bq},
\nonumber \\
 \,  [b_{\alpha,\bk},b_{\beta,\bq}] &=& [b^\dagger_{\alpha,\bk},b^\dagger_{\beta,\bq}] =0.
\label{commutator}
\end{eqnarray}
Here, the $g_\alpha(\bp,\bq)$ functions are given by Eq.~\eqref{g-func-hall} 
and 
\begin{equation}
   F^2_{\alpha,\bq} =  \sum_\bp g_\alpha(\bp,\bq) g_\alpha(\bp-\bq,-\bq).
\label{f-func-hall01}
\end{equation}
Interestingly, it is possible to show that the $F_{\alpha,\bq}$ functions can be
explicitly written in terms of the $\hat{B}_{i,\bp}$ coefficients,
see Eq.~\eqref{f-func-hall}. 
It is worth emphasizing that the boson operators $b_\alpha$
are defined with respect to the reference state $|{\rm FM}\rangle$.
 
Once the boson operators $b_\alpha$ are defined, we can derive the
bosonic representation of any operator $\mathcal{O}$ that is expanded
in terms of the fermion operators $c$.    
As discussed in Ref.~\onlinecite{doretto05}, such a procedure consists
of calculating the commutators $[\mathcal{O}, b^\dagger_{\alpha,\bq}]$,
writing them in terms of the boson operators $b_\alpha$, and
determining the action of the operator $\mathcal{O}$ in the reference
state $|{\rm FM}\rangle$. 
For instance, let us consider the projected electron density operator 
$\bar{\rho}_{a\,\uparrow}(\bk)$ [Eq.~\eqref{proj-dens-op}]. 
From Eqs.~\eqref{proj-dens-op}, \eqref{b-op}, and \eqref{comut-s-rho}, 
one finds that   
\begin{equation}
[\bar{\rho}_{a\,\uparrow}(\bk), b^\dagger_{\alpha,\bq}] =
     -\sum_\bp\frac{G_a(\bp,\bk)}{F_{\alpha,\bq} } g_\alpha(\bp-\bk,\bq)
     c_{\bp-\bk-\bq\,\downarrow}c_{\bp\,\uparrow}.
\label{commutator01}
\end{equation} 
Differently from the 2DEG at $\nu=1$,\cite{doretto05} it is not
possible to express the commutator \eqref{commutator01} in terms of
the boson operators $b_\alpha$ and therefore, it is not easy to
determine the expansion of $\bar{\rho}_{a\,\uparrow}(\bk)$ in terms of the bosons
$b_\alpha$ that satisfies the commutator \eqref{commutator01}. 
This is related to the fact that for the Chern insulators the algebra 
of the spin and electron density operators is not closed (see
discussion above).  
We then proceed as follows: In principle, we can write
\begin{equation}
 c^\dagger_{\bp-\bq\,\downarrow}c_{\bq\,\uparrow} =
  \sum_\alpha\sum_\bk H_\alpha(\bk,\bp,\bq)b^\dagger_{\alpha,\bk},
\label{cc1}
\end{equation}
where the $H_\alpha(\bk,\bp,\bq)$ function satisfies the relation
[compare Eqs.~\eqref{cc1} and \eqref{b-op}] 
\begin{equation}
 \sum_\bp g_\alpha(\bp,\bq)H_\beta(\bk,\bp,\bq) = 
          \delta_{\alpha,\beta}\delta_{\bk,\bq}F_{\alpha,\bq}.
\label{h-func4}
\end{equation}
With the aid of Eq.~\eqref{cc1}, the commutator \eqref{commutator01}
reads 
\begin{eqnarray}
 [\bar{\rho}_{a\,\uparrow}(\bk), b^\dagger_{\alpha,\bq}] &=& 
     -\frac{1}{F_{\alpha,\bq}}\sum_\beta\sum_{\bk',\bp} 
     G_a(\bp,\bk)g_\alpha(\bp-\bk,\bq)
\nonumber \\
 && \,\times H_\beta(\bk',\bp,\bk+\bq)b^\dagger_{\beta,\bk'}.
\label{commutator02} 
\end{eqnarray} 
It is then easy to see that the expansion 
\begin{eqnarray}
  \bar{\rho}_{a\,\uparrow}(\bk) &=& 
    -\sum_{\alpha,\beta}\sum_{\bk',\bp,\bq}  \frac{1}{F_{\alpha,\bq}}
     G_a(\bp,\bk)g_\alpha(\bp-\bk,\bq)
\nonumber \\
 &&\,\times H_\beta(\bk',\bp,\bk+\bq)b^\dagger_{\beta,\bk'} b^\dagger_{\alpha,\bq}
\label{rho-boso01}
\end{eqnarray} 
of $\bar{\rho}_{a\,\uparrow}(\bk)$ in terms of the bosons $b_\alpha$ 
satisfies the commutator \eqref{commutator02}. Apart from a
constant (see below), Eq.~\eqref{rho-boso01} corresponds to 
the bosonic representation of the electron density operator
$\bar{\rho}_{a\,\uparrow}(\bk)$.  

Although it is difficult to solve Eq.~\eqref{h-func4} and calculate
$H_\alpha(\bk,\bp,\bq)$, it is indeed possible to determine 
the particular value for $\bk = \bq$, $H_\alpha(\bq,\bp,\bq)$. 
Comparing Eqs.~\eqref{f-func-hall01} and \eqref{h-func4}, we see that
\begin{equation}
  H_\alpha(\bq,\bp,\bq) =
  \frac{1}{F_{\alpha,\bq}}g_\alpha(\bp-\bq,-\bq). 
\label{h-func-approx}
\end{equation} 
Note that by keeping only the term $\bk' = \bk+\bq$ in the momentum
sum in Eq.~\eqref{rho-boso01} and using Eq.~\eqref{h-func-approx}, an
approximated bosonic representation for the electron density operator
$\bar{\rho}_{a\,\uparrow}(\bk)$ follows.
Similar considerations hold for $\bar{\rho}_{a\,\downarrow}(\bk)$. 
We then arrive at the following {\it approximated} bosonic representation for the
electron density operator:  
\begin{equation}
\bar{\rho}_{a\, \sigma}(\bk) \approx 
         \frac{1}{2}N\delta_{\sigma,\uparrow}\delta_{\bk,0}
         + \sum_{\alpha,\beta}\sum_\bq 
         \gm_{\alpha\beta\,a\,\sigma}(\bk,\bq)b^\dagger_{\beta,\bk+\bq}b_{\alpha,\bq}.
\label{rho-boso02}
\end{equation}
Here, the first term is related to the action of $\bar{\rho}_{a\,\sigma}(\bk) $ 
in the reference state $|{\rm FM}\rangle$ and the
$\gm_{\alpha\beta\,a\,\sigma}(x,y)$ function is given by 
Eq.~\eqref{galphabeta-hall01} in Appendix \ref{ap:boso}. 
Similar to the $F_{\alpha,\bq}$ function \eqref{f-func-hall01},
the $\gm_{\alpha\beta\,a\,\sigma}(x,y)$ function can also be expressed in
terms of the $\hat{B}_{i,\bp}$ coefficients [see Eq.~\eqref{galphabeta-hall02}].

In summary, we show that the bosonization formalism introduce in
Ref.~\onlinecite{doretto05} for the 2DEG at $\nu=1$ can be extended to
flat-band Chern insulators with a half-filled energy band. In both
cases, it is possible to show that the particle-hole pair excitations
can be approximately treated as bosons as long as the number of such
excitations is small. While for the 2DEG an exact bosonic
representation for the electron density operator can be derived, here,
due to the fact that the algebra of the spin and electron density operators
is not closed, only an approximated bosonic
representation for the electron density operator can be obtained.
We should note that the approximated bosonic representation for
the electron density operator \eqref{rho-boso02} is similar to
the {\it exact} one derived for the 2DEG at $\nu=1$ [see Eq.~(27) from
Ref.~\onlinecite{doretto05}].

%%%%%%%%%%%%%%%%%%%%%%%%%%%%%%%%%%%%%%%%%%%%%%%%%%%%%%%%%%%%%%%%%%%%%%%%%%%%%%%%%%%%%
\subsection{Topological Hubbard model I}
\label{sec:boso-model}

Let us now consider a square lattice Hubbard model at $1/4$-filling
whose Hamiltonian is given by
\begin{equation}
 \mathcal{H}_{Ch} = \mathcal{H}_0 + \mathcal{H}_U. 
\label{ham}
\end{equation}
Here, $\mathcal{H}_0$ is the tight-binding model \eqref{ham-0} with
$t_2 = t_1/\sqrt{2}$ (nearly flat-band limit) and
\begin{equation}
 \mathcal{H}_U = U\sum_{i}\sum_{a = A,B}
                \hat{\rho}_{i\, a\, \uparrow}\hat{\rho}_{i\, a\, \downarrow}
\label{ham-u}
\end{equation}
is the one-site Hubbard term with  
$\hat{\rho}_{i\, a\,\uparrow}$ being the electron density operator
\eqref{density-op} and $U > 0$. 
In momentum space, $\mathcal{H}_U$ reads  
\begin{equation}
 \mathcal{H}_U = \frac{U}{N}\sum_a\sum_\bk 
           \hat{\rho}_{a \uparrow}(-\bk)\hat{\rho}_{a \downarrow}(\bk)
\label{ham-u-k}
\end{equation}
with $\hat{\rho}_{a \sigma}(\bk)$ given by Eq.~\eqref{density-op-2}
and $N$ being the number of unit cells, as mentioned before. 
Since the choice $t_2 = t_1/\sqrt{2}$ implies that the energy bands of
$\mathcal{H}_0$ have nonzero Chern numbers [Eq.~\eqref{chern-number2}],
the Hamiltonian \eqref{ham} corresponds to a topological Hubbard model.
In the following, we apply the bosonization formalism introduced in the previous
section to study the flat-band ferromagnetic phase of the correlated
Chern insulator \eqref{ham}.

We start by projecting the Hamiltonian \eqref{ham} into the lowest
energy $c$ bands:
\begin{equation}
 \mathcal{H}_{Ch} \rightarrow 
 \mathcal{\bar{H}}_{Ch} = \mathcal{\bar{H}}_0 + \mathcal{\bar{H}}_U.
\label{ham-proj}
\end{equation}
Here
\begin{equation}
 \mathcal{\bar{H}}_0 =  \sum_\bq 
    \omega_{c,\bq} c^\dagger_{\bq\,\sigma}c_{\bq\,\sigma} 
\label{ham-0-proj}
\end{equation}
[see Eq.~\eqref{diag-ham-0}] and $\mathcal{\bar{H}}_U$ is given by Eq.~\eqref{ham-u-k} with the replacement
$\hat{\rho}_{a\, \sigma}(\bk) \rightarrow  \bar{\rho}_{a\,\sigma}(\bk)$.
Some issues about the relevant energy scales need to be emphasized:
On the one hand, in order for the projection scheme to the lowest $c$ bands to remain
valid, the energy scale $U$ must be smaller than the energy separation
$\Delta$ between the bands $c$ and $d$, see Fig.~\ref{fig:gap}. 
Otherwise, the on-site interaction would mix the different bands and 
it would no longer be valid to characterize them in terms of Chern numbers associated 
with the non-interacting model. On the other hand, we consider the on-site interaction $U$ to
be much larger than the bandwidth of the (almost flat) $c$ bands, such that their
dispersion may be neglected in the remainder of the section. In this sense, the $c$ bands
are reminiscent of the highly-degenerate flat Landau levels of a 2DEG in a strong
magnetic field. 

Following the same procedure used to determine the bosonic representation 
of the projected electron density operator \eqref{proj-dens-op}, we show that
the bosonic representation of $\mathcal{\bar{H}}_0$ is simply a
constant $E_0$ 
[recall that in the flat-band limit, the kinetic energy is quenched, see
Eqs.~\eqref{boso-aux1}-\eqref{boso-aux2} for details]. The bosonic representation of the
projected on-site Hubbard term $\mathcal{\bar{H}}_U$ can be easily derived by
substituting  Eq.~\eqref{rho-boso02} into $\mathcal{\bar{H}}_U$  and normal
ordering the resulting expression. We then arrive at the boson model
\begin{equation}
\mathcal{H}_B = E_0 + \mathcal{H}^{(2)}_B +  \mathcal{H}^{(4)}_B, 
\label{h-boso}
\end{equation}
where $E_0 = 2.44N$, the quadratic boson term is given by 
\begin{equation}
 \mathcal{H}^{(2)}_B = \sum_{\alpha,\beta}\,\sum_{\bp \in BZ} 
               \epsilon^{\alpha\beta}_\bp \,
                b^\dagger_{\beta,\bp} b_{\alpha,\bp},
\label{h-boso2}
\end{equation}
and the boson-boson interaction term reads
\begin{equation}
 \mathcal{H}^{(4)}_B = \frac{1}{N}\sum_{\alpha,\beta,\alpha',\beta'}\sum_{\bk,\bq,\bp}
       V^{\alpha\beta\alpha'\beta'}_{\bk,\bq,\bp} 
       b^\dagger_{\beta',\bp+\bk}b^\dagger_{\beta,\bq-\bk}
       b_{\alpha,\bq}b_{\alpha',\bp}.
\label{h-boso4}
\end{equation}
Here 
\begin{eqnarray}
 \epsilon^{\alpha\beta}_\bp &=& 
   \frac{U}{2}\sum_a \gm_{\alpha\,\beta\,a\,\downarrow}(0,\bp)  
\nonumber \\
   &+&  \frac{U}{N}\sum_{a,\alpha',\bk} 
   \gm^*_{\beta\,\alpha'\,a\,\uparrow}(\bk,\bp)
   \gm_{\alpha\,\alpha'\,a\,\downarrow}(\bk,\bp), 
\label{ep-b-model} 
\\
 V^{\alpha\beta\alpha'\beta'}_{\bk,\bq,\bp} &=& 
  \frac{U}{N}\sum_a 
   \gm^*_{\alpha\,\beta\,a\,\uparrow}(-\bk,\bq)
   \gm_{\alpha'\,\beta'\,a\,\downarrow}(\bk,\bp), 
\label{int-b-model}
\end{eqnarray}
with the $\gm_{\alpha\,\beta\,a\,\sigma}(\bk,\bp)$ function given
by Eq.~\eqref{galphabeta-hall01}
and $\alpha, \beta, \alpha', \beta'  = 0,1$.
Therefore, within the bosonization scheme introduced in
Sec.~\ref{sec:boso}, the Hubbard model \eqref{ham} is mapped into the
effective {\it interacting} boson model \eqref{h-boso}. 

In order to discuss some features of the effective boson model
\eqref{h-boso}, let us first neglected the quartic boson term 
$\mathcal{H}^{(4)}_B$ and consider only   
\begin{equation}
  \mathcal{H}_B \approx E_0 + \mathcal{H}^{(2)}_B. 
\label{h-boso-harmonic}
\end{equation}
Such a lowest-order approximation is the so-called harmonic
approximation. The quadratic Hamiltonian \eqref{h-boso-harmonic} 
can be easily diagonalized, namely 
\begin{equation}
\mathcal{H}_B = E_0 + \sum_{\mu = \pm}\,\sum_{\bp \in BZ} 
               \Omega_{\mu,\bp} 
                a^\dagger_{\mu,\bp} a_{\mu,\bp},
\label{ham-harm}
\end{equation}
where the dispersion relations of the bosons $a_\pm$  read
\begin{equation}
\Omega_{\pm,\bp} = \pm\Delta_\bp + \sqrt{\epsilon^2_\bp 
                - \epsilon^{10}_\bp\epsilon^{01}_\bp}
\label{sw-chern}
\end{equation}
with
\begin{equation}
 \epsilon_\bp = \frac{1}{2}[\epsilon^{00}_\bp + \epsilon^{11}_\bp],  
\;\;\;\; \;\;\;\;\;\;\;\;
 \Delta_\bp = \frac{1}{2}[\epsilon^{00}_\bp  - \epsilon^{11}_\bp],  
\label{sw-chern-aux}
\end{equation}
and the $\epsilon^{\alpha\beta}_\bp$ given by Eq.~\eqref{ep-b-model}.

The ground state of the Hamiltonian \eqref{ham-harm} is the vacuum
state for the bosons $a_\mu$. Since 
$b_{\alpha,\bk}|{\rm FM}\rangle = a_{\mu,\bk}|{\rm FM}\rangle = 0$,
the ground state of \eqref{ham-harm} is indeed the spin polarized ferromagnet
state \eqref{fm}. This result indicates that the topological Hubbard
model \eqref{ham} has a stable flat-band ferromagnetic phase.

\begin{figure}[t]
\centerline{\includegraphics[width=9.5cm]{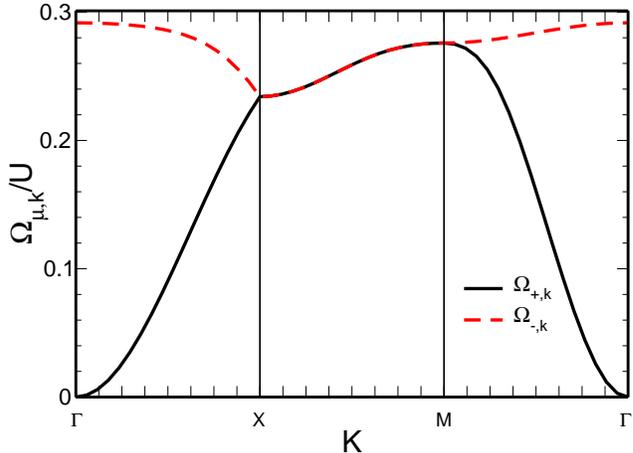} }
\caption{(Color online) Dispersion relation \eqref{sw-chern} of the
  elementary excitations of the effective boson model \eqref{h-boso}
  along paths in the Brillouin zone [Fig.~\ref{fig:lattice}(c)] at the
  harmonic approximation. Such a spectrum corresponds to the spin wave
  excitations of the flat-band ferromagnetic phase of the Chern
  insulator \eqref{ham}.     
}
\label{fig:disp-u}
\end{figure}

The stability of the ferromagnetic ground state is also 
corroborated by the dispersion relations $\Omega_{\pm,\bp}$
for the bosons $a_\pm$ (Fig.~\ref{fig:disp-u}), which correspond to
the spin-wave spectrum of the 
flat-band ferromagnetic ground state $|{\rm FM}\rangle$.  
One notices that the energy scale of the excitations is given by the on-site
repulsion energy $U$ instead of the nearest-neighbor hopping energy
$t_1$ since the boson representation of $\mathcal{\bar{H}}_0$ is simply a
constant $E_0$.      
The excitation spectrum has two branches, a gapless one
($\Omega_{+,\bp}$), with the 
Goldstone mode at the centre of the Brillouin zone, and a gapped
branch ($\Omega_{-,\bp}$), with the lowest energy excitation 
at the $X$ point [see Fig.~\ref{fig:lattice}(c)]. The presence of the
Goldstone mode indicates the breaking of a continuous SU(2) symmetry
as expected for the correlated Chern insulator \eqref{ham}, 
see Sec.~\ref{sec:symmetries} for details.   
Finally, it should be mentioning that the spectrum shown in Fig.~\ref{fig:disp-u} is
qualitatively similar to the spin-wave excitation spectrum of the two-dimensional
Mielkes model (flat-band limit) derived by Kusakabe and Aoki in the
weak-coupling regime [see Fig.~1(a) from Ref.~\onlinecite{aoki94}].

%%%%%%%%%%%%%%%%%%%%%%%%%%%%%%%%%%%%%%%%%%%%%%%%%%%%%%%%%%%%%%%%%%%%%%%%%%%%%%%%%%%%%
\section{Ferromagnetism in a flat-band Z$_2$ topological insulator}
\label{sec:shall}

In this section, we study the flat-band ferromagnetic phase of a
topological Hubbard model that preserves time-reversal symmetry. Due
to the similarities with Sec.~\ref{sec:hall}, here we just quote the
main results and comment on the differences between flat-band Chern and
topological insulators.

%%%%%%%%%%%%%%%%%%%%%%%%%%%%%%%%%%%%%%%%%%%%%%%%%%%%%%%%%%%%%%%%%%%%%%%%%%%%%%%%%%%%%  
\subsection{Time-reversal symmetric square lattice $\pi$-flux model}
\label{sec:model02}

Similar to Sec.~\ref{sec:model0}, we consider $N$ spinfull
noninteracting electrons hopping on a bipartite square lattice where
each sublattice, $A$ and $B$, has $N_A = N_B = N$ sites.   
The Hamiltonian of the system $\mathcal{H}^{TRS}_0$ is given by
Eq.~\eqref{ham-0}  but now we assume that 
$\gamma(\uparrow) = -\gamma(\downarrow) = 1$ for the
nearest-neighbor hopping energy \eqref{hopping1}. Such a choice
implies that time-reversal symmetry is preserved, see 
Appendix~\ref{ap:symmetries} for details.  
$\mathcal{H}^{TRS}_0$ can be seen as two copies of the
spinless $\pi$-flux model\cite{neupert11} where electrons with spin $\uparrow$ and
$\downarrow$ are in the presence of opposite (fictitious) 
staggered $\pm\pi$ flux patterns, see Fig.~\ref{fig:lattice}(a).   

In momentum space, $\mathcal{H}^{TRS}_0$ assumes the form 
\eqref{ham-0-k} but with 
$h^\downarrow_\bk = h^{\uparrow\;*}_{-\bk}$, which is precisely a consequence of
invariance under time-reversal transformations.
It is easy to see that $\mathcal{H}^{TRS}_0$ can be diagonalized by
the the Bogoliubov transformation
\begin{eqnarray}
 c^\dagger_{\bk\,A\,\uparrow} &=& u_\bk d^\dagger_{\bk\,\uparrow} 
                     + v^*_\bk c^\dagger_{\bk\,\uparrow},
%\nonumber \\
\;\;\;\;\;\;
 c^\dagger_{\bk\,B\,\uparrow} = v_\bk d^\dagger_{\bk\,\uparrow} 
                     - u^*_\bk c_{\bk\,\uparrow},
\nonumber \\
&& \label{bogo-qhe-sqhe} \\
c^\dagger_{\bk\,A\,\downarrow} &=& u^*_{-\bk} d^\dagger_{\bk\,\downarrow} 
                             + v_{-\bk} c^\dagger_{\bk\,\downarrow},
%\nonumber \\
\;\;\;
 c^\dagger_{\bk\,B\,\downarrow}  =  v^*_{-\bk} d^\dagger_ {\bk\,\downarrow}
                               - u_{-\bk} c^\dagger_{\bk\,\downarrow},
\nonumber
\end{eqnarray}
with the Bogoliubov coefficients $u_{\bk}$ and $v_{\bk}$ given by
Eq.~\eqref{coef-bogo}. 
Note that since $h^\downarrow_\bk = h^{\uparrow\;*}_{-\bk}$, the
canonical transformation for spin $\uparrow$ electrons differs
from the one  for spin $\downarrow$ electrons in contrast to the
Chern insulator discussed in Sec.~\ref{sec:model0}, where
both transformations are equal [see Eq.~\eqref{bogo-qhe}].  
After diagonalization, the Hamiltonian also assumes the form
\eqref{diag-ham-0}, but now the  $4\times 4$ matrix $\mathcal{H'}_\bk$ reads
\begin{equation}
\mathcal{H'}_\bk = \left( 
  \begin{array}{cc} h'_\bk & 0 \\
                              0 & h'_{-\bk} 
  \end{array} \right)
\end{equation}
with the $2\times 2$ diagonal matrix $h'_\bk$ giving by Eq.~\eqref{h-matrix}. 
Similarly to the Chern insulator discussed in Sec.~\ref{sec:model0}, the
resulting band structure comprises two doubly degenerate bands
$c$ and $d$, whose dispersion relations $\omega_{c/d,\sigma,\bk}$ are
also given by Eq.~\eqref{omega-cd} [see Fig.~\ref{fig:bands}]. 
As required by time-reversal symmetry,\cite{goerbig12}
$\omega_{c/d,\sigma,\bk} = \omega_{c/d,-\sigma,-\bk}$. 
In particular, for the $\pi$-flux model $\mathcal{H}^{TRS}_0$, %we have
$\omega_{c/d,\uparrow,\bk} = \omega_{c/d,\downarrow,\bk}$.  

Again, the $c$ and $d$ bands are topologically nontrivial. Indeed, 
it follows from Eq.~\eqref{chern-number2} in combination with the fact
that $\gamma(\uparrow) = -\gamma(\downarrow) = 1$ that
the Chern numbers of the $c$ and $d$ bands 
are given by\cite{neupert12}
\[
  C^d_\uparrow = - C^d_\downarrow = -1
 \;\;\;\;\;\;\;  {\rm and} \;\;\;\;\;\;\; 
  C^c_\uparrow = - C^c_\downarrow = +1,
\]
i.e., $C^{c/d}_\sigma = - C^{c/d}_{-\sigma}$ as required by time--reversal symmetry. 
As a consequence,  the {\it charge} Chern numbers\cite{neupert12}
of the $c$ and $d$ bands vanish, 
\[
 C^{c/d}_c = \frac{1}{2}\left( C^{c/d}_\uparrow + C^{c/d}_\downarrow \right) = 0,
\]
while the corresponding {\it spin} Chern numbers are nonzero,
\[
 C^{c/d}_s = \frac{1}{2}\left( C^{c/d}_\uparrow - C^{c/d}_\downarrow \right) 
 =  \pm 1.
\]
Since $\mathcal{H}^{TRS}_0$ conserves the $\hat{z}$-component of the total
spin (see Sec.~\ref{sec:symmetries}), the Z$_2$ topological
invariant for the $c$ and $d$ bands are simply given by\cite{kane13,sheng06} 
\begin{equation}
  \nu_{c/d} = C^{c/d}_s\;{\rm mod}\; 2 = 1,
\end{equation}
implying that $\mathcal{H}^{TRS}_0$ is a Z$_2$ topological insulator.
Indeed, at half-filling (configuration not considered here),
$\mathcal{H}^{TRS}_0$ should display a QSHE 
with the {\it spin} Hall conductivity $\sigma^{SH}_{xy} = eC_s/2\pi$.

%%%%%%%%%%%%%%%%%%%%%%%%%%%%%%%%%%%%%%%%%%%%%%%%%%%%%%%%%%%%%%%%%%%%%%%%%%%%%%%%%%%%%
\subsection{Bosonization scheme for a flat-band topological insulator}
\label{sec:boso-sqhe}

In this section, we introduce a bosonization scheme for a flat-band 
Z$_2$ topological insulator similar to the one for the flat-band
Chern insulator discussed in Sec.~\ref{sec:boso}. 
As shown below, the two bosonization schemes are quite similar, but
there are important differences due to the fact that here time-reversal symmetry
is preserved. Again, we focus on the nearly flat-band limit of the
tight-binding model $\mathcal{H}^{TRS}_0$ ($t_2 = t_1/\sqrt{2}$) 
at $1/4$-filling ($N_A = N_B = N$). We restrict the Hilbert space to
the lowest energy $c$ bands and also assume that the ground 
state of $\mathcal{H}^{TRS}_0$ is given by the ferromagnet state \eqref{fm}.  

Instead of the spin operator \eqref{spin-op} at site $i$ of the sublattice $a$,
we now consider the following spin operator  
\begin{equation}
\bS_{i\,a b} = \frac{1}{2}c^\dagger_{i\, a\,\sigma}\hat{\sigma}_{\sigma\,\sigma'}
                   c_{i\, b\,\sigma'}
\label{spin-op2}
\end{equation}
with $(a,b) = (A,B)$ and $(B,A)$. Using Eqs.~\eqref{fourier} and
\eqref{fourier-spin}, the expression of the spin operators
\eqref{spin-op2} in momentum space can be derived. In particular, 
\begin{eqnarray}
 S^+_{\bq,a b} &=& \sum_\bp c^\dagger_{\bp-\bq\, a\,\uparrow}c_{\bp\, b\,\downarrow},
\nonumber \\
 S^-_{\bq,a b} &=& \sum_\bp c^\dagger_{\bp-\bq\, a\,\downarrow}c_{\bp\, b\,\uparrow},
\end{eqnarray}
where $S^\pm_{\bq, ab} = S^x_{\bq,ab} \pm iS^y_{\bq,ab}$. The spin
operators $\bar{S}^\pm_{\bq,ab}$  projected into the $c$ bands, i.e.,
the equivalent of Eq.~\eqref{aux-spin-op}, now read 
\begin{eqnarray}
 \bar{S}^+_{\bq,ab} &=& \sum_\bp
           G^*_{ab}(-\bp,-\bq)c^\dagger_{\bp-\bq\,\uparrow}c_{\bp\,\downarrow},
\nonumber \\
 \bar{S}^-_{\bq,ab} &=& \sum_\bp
          G_{ab}(\bp,\bq)c^\dagger_{\bp-\bq\,\downarrow}c_{\bp\,\uparrow},
\label{aux-spin-op2}
\end{eqnarray}
where 
\[
  G_{AB}(\bp,\bq) =- v_{-\bp+\bq}u_\bp 
 \;\;\;\; {\rm and}  \;\;\;\;
  G_{BA}(\bp,\bq) = -u_{-\bp+\bq}v_\bp. 
\]
Again, we consider the following linear combination of the spin operators
\begin{equation}
 \bar{S}^\lambda_{\bq,\alpha}  = \bar{S}^\lambda_{\bq,AB} 
                             + (-1)^\alpha\bar{S}^\lambda_{\bq,BA}
\label{spin-alpha-op2}
\end{equation}
with $\lambda = x,y,z$ and $\alpha=0,1$. In particular, we have
\begin{eqnarray}
 \bar{S}^+_{\bq,\alpha} &=& \sum_\bp 
          g^*_\alpha(-\bp,-\bq)c^\dagger_{\bp-\bq\,\uparrow}c_{\bp\,\downarrow},
\nonumber \\
 \bar{S}^-_{\bq,\alpha} &=& \sum_\bp
          g_\alpha(\bp,\bq)c^\dagger_{\bp-\bq\,\downarrow}c_{\bp\,\uparrow},
\label{spin-alpha-pm-op2}
\end{eqnarray}
where the $g_\alpha(\bp,\bq)$ functions are now defined by
\begin{equation}
  g_\alpha(\bp,\bq) = -v_{-\bp+\bq}u_\bp - (-1)^\alpha u_{-\bp+\bq}v_\bp
\label{g-func-shall}
\end{equation}
with $\alpha = 0,1$ [compare Eqs.~\eqref{g-func-hall} and \eqref{g-func-shall}].

Similar to the flat-band Chern insulators, the algebra of the
projected spin and electron density operators is not closed. For
instance, the equivalent of the commutator \eqref{comut-ss} 
now reads 
\begin{eqnarray}
[\bar{S}^+_{\bq,\alpha}, \bar{S}^-_{\bq',\beta}] &=& \sum_\bp  
           \left[ g^*_\alpha(-\bp+\bq',-\bq)g_\beta(\bp,\bq')
                   c^\dagger_{\bp-\bq-\bq'\,\uparrow}c_{\bp\,\uparrow}
           \right.
\nonumber \\
&-&  \left.    g^*_\alpha(-\bp,-\bq)g_\beta(\bp-\bq,\bq')
             c^\dagger_{\bp-\bq-\bq'\,\downarrow}c_{\bp\,\downarrow}
           \right]. \;
\label{comut-ss-z2} 
\end{eqnarray}
The complete algebra of the projected spin and electron density operators can
be found in Appendix~\ref{ap:boso-spin}.  

The construction of the boson operators $b_\alpha$ in terms of the
fermion operators $c$ follows the same procedure outlined in
Eqs.~\eqref{excitation-chern}--\eqref{b-op}. 
Importantly, the spin operator that defines the particle-hole
excitation [Eq.~\eqref{excitation-chern}] is now given by
Eq.~\eqref{spin-alpha-op2}. Again, we can define two
sets of independent boson operators, namely
\begin{eqnarray}
b_{\alpha,\bq} &=& \frac{1}{F_{\alpha,\bq}}
  \sum_\bp g^*_\alpha(-\bp,\bq)c^\dagger_{\bp+\bq\,\uparrow}c_{\bq\,\downarrow},
\nonumber\\
&& \nonumber \\
b^\dagger_{\alpha,\bq} &=& \frac{1}{F_{\alpha,\bq}}
  \sum_\bp g_\alpha(\bp,\bq)c^\dagger_{\bp-\bq\,\downarrow}c_{\bq\,\uparrow},
\label{b-op-shall}
\end{eqnarray}
with $\alpha = 0,1$, that obey the commutation relations \eqref{commutator}.
Here, the $g_\alpha(\bp,\bq)$ functions are given by
Eq.~\eqref{g-func-shall} and  
\begin{eqnarray}
 F^2_{\alpha,\bq} &=& \sum_\bp g^*_\alpha(-\bp+\bq,\bq)g_\alpha(\bp,\bq)
\label{f-func-sqhe}
\end{eqnarray}
[see Eq~\eqref{f-func-sqhe2} for the expression of the
$F_{\alpha,\bq}$ function in terms of the coefficients $\bhat_{i,\bp}$].
It is worth mentioning that although for both Chern and Z$_2$
topological insulators the bosons $b_\alpha$ are linear combinations of
particle-hole pair excitations, for the former the particle and the
hole are on the same sublattice [see Eqs.~\eqref{spin-alpha-op} and
\eqref{b-op}] while, for the latter, the particle and the hole are on
different sublattices [see Eqs.~\eqref{spin-alpha-op2} and
\eqref{b-op-shall}].\cite{kumar14}      

Finally, the electron density operator \eqref{density-op-2}
projected into the $c$ bands [the equivalent of
Eq.~\eqref{proj-dens-op}] now reads  
\begin{equation}
 \bar{\rho}_{a\, \sigma}(\bk) = \sum_\bp 
    G_{a\,\sigma}(\bp,\bk)c^\dagger_{\bp-\bk\,\sigma}c_{\bp\,\sigma} 
\label{density-op-proj2}
\end{equation}
with 
\begin{eqnarray}
  G_{A\,\uparrow}(\bp,\bk) &=& v^*_{\bp-\bk}v_\bp,   
 \;\;\;\;\;\;\;\;
  G_{B\,\uparrow}(\bp,\bk) = u^*_{\bp-\bk}u_\bp, 
\nonumber \\
&& \label{gabsigma-functions} \\
  G_{A\,\downarrow}(\bp,\bk) &=& v_{-\bp+\bk}v^*_{-\bp},   
 \;\;\;\; 
  G_{B\,\downarrow}(\bp,\bk) = u_{-\bp+\bk}u^*_{-\bp}.
\nonumber 
\end{eqnarray}
Following the procedure outlined in Eqs.~\eqref{commutator01}--\eqref{rho-boso02}, 
the bosonic representation of \eqref{density-op-proj2} can be easily
derived. It is also given by Eq.~\eqref{rho-boso02} but with the
$\gm_{\alpha\beta\,a\,\sigma}(x,y)$ 
function now given by Eq.~\eqref{galphabeta-shall01}. Here, the
equivalent of Eq.~\eqref{h-func-approx}  reads 
\begin{equation}
  H_\alpha(\bq,\bp,\bq) =
  \frac{1}{F_{\alpha,\bq}}g^*_\alpha(-\bp+\bq,\bq). 
\label{h-func-approx2}
\end{equation}

%%%%%%%%%%%%%%%%%%%%%%%%%%%%%%%%%%%%%%%%%%%%%%%%%%%%%%%%%%%%%%%%%%%%%%%%%%%%%%%%%%%%%
\subsection{Topological Hubbard model II}
\label{sec:boso-model02}

In this section, we consider a correlated topological insulator on a
square lattice described by the Hamiltonian
\begin{equation}
 \mathcal{H}_{Z2} = \mathcal{H}^{TRS}_0 + \mathcal{H}_U, 
\label{ham-z2}
\end{equation}
where $\mathcal{H}^{TRS}_0$ is the square lattice $\pi$-flux model
discussed in Sec.~\ref{sec:model02} in the nearly flat-band limit 
($t_2 = t_1/\sqrt{2}$) and $\mathcal{H}_U$ is the repulsive on-site
Hubbard term \eqref{ham-u}. Again,  $1/4$-filling is assumed.
The topological Hubbard model \eqref{ham-z2} has recently been discussed
by Neupert {\it et al.}\cite{neupert12} Similarly to Sec~\ref{sec:boso},
we now apply the bosonization formalism introduced in
Sec.~\ref{sec:boso-sqhe} to study the flat-band ferromagnetic phase of
the Hamiltonian \eqref{ham-z2}.

\begin{figure}[t]
\centerline{\includegraphics[width=9.5cm]{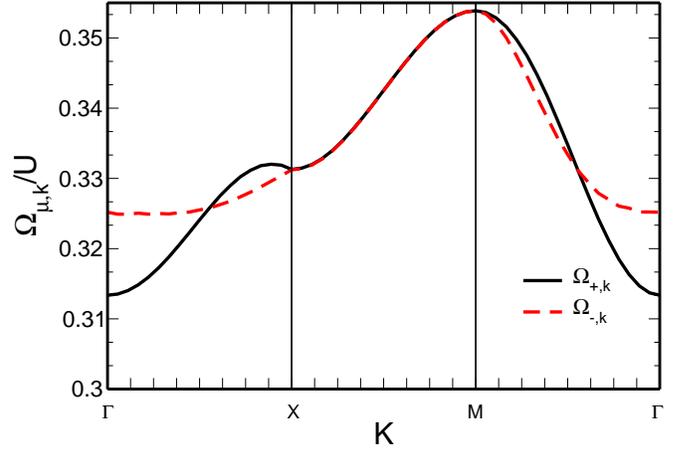}}
\caption{(Color online) Dispersion relation \eqref{sw-z2} of the
  elementary excitations of the effective boson model $\mathcal{H}^{Z2}_B$ 
  along paths in the Brillouin zone [Fig.~\ref{fig:lattice}(c)] at the
  harmonic approximation. Such a spectrum corresponds to the spin wave
  excitations of the flat-band ferromagnetic phase of the
  Z$_2$ topological insulator \eqref{ham-z2}. 
}
\label{fig:disp-u-sqhe}
\end{figure}

Following the lines of Sec.~\ref{sec:boso-model}, the first step is to
project the Hamiltonian \eqref{ham-z2} into the lowest-energy $c$ bands, i.e, 
\[
   \mathcal{H}_{Z2} \rightarrow \mathcal{\bar{H}}_{Z2}.
\]
We then map the projected Hubbard model $\mathcal{\bar{H}}_{Z2}$ to an
effective interacting boson model $\mathcal{H}^{Z2}_B$. We find that   
$\mathcal{H}^{Z2}_B$ has the same form as the boson Hamiltonian
\eqref{h-boso} but with the $\gm_{\alpha\beta\,a\,\sigma}(x,y)$
function given by Eq.~\eqref{galphabeta-shall01}.

Within the harmonic approximation, the effective boson model
$\mathcal{H}^{Z2}_B$ can be diagonalized and it assumes the
form \eqref{ham-harm}. The dispersion relation of the bosons $a_\pm$
is equal to \eqref{sw-chern}, i.e.,
\begin{equation}
\Omega^{Z2}_{\pm,\bp} = \pm\Delta_\bp + \sqrt{\epsilon^2_\bp 
                - \epsilon^{10}_\bp\epsilon^{01}_\bp}
\label{sw-z2}
\end{equation}
but with the $\gm_{\alpha\beta\,a\,\sigma}(x,y)$ function given by 
Eq.~\eqref{galphabeta-shall01}.

Similarly to the correlated Chern insulator discussed in
Sec.~\ref{sec:boso-model}, the ground state of $\mathcal{H}^{Z2}_B$ is
the vacuum state for the bosons $a_\pm$, i.e., the ferromagnetic state 
$|{\rm FM}\rangle$ [Eq.~\eqref{fm}], and
the excitation spectrum [the dispersion relation \eqref{sw-z2} of the
bosons $a_\pm$]  of such a flat-band ferromagnet has two branches, 
see Fig.~\ref{fig:disp-u-sqhe}. 
However, in contrast to the spin-wave spectrum of a Chern insulator
(Fig.~\ref{fig:disp-u}), here both branches are gapped at zero wave
vector.   
Differently from the flat-band Chern insulator
\eqref{ham},  where a continuous SU(2) symmetry is
broken, the ground state of the correlated topological insulator
preserves the U(1) spin rotation symmetry of the 
Hamiltonian \eqref{ham-z2} 
(see Sec.~\ref{sec:symmetries} for more details).
As a consequence, a Goldstone mode is absent in the excitation
spectrum.   
Again, the energy scale of the spin-wave excitations is given
by the on-site repulsion energy $U$. 
The excitation gap $\Delta = \Omega^{Z2}_{+,\bk=0} = 0.3134\,U$, 
which is at the centre of the Brillouin zone, agrees with the exact
diagonalization data from Neupert {\it et al.}, $\Delta  \approx
0.30\,U$.\cite{neupert12} 
The fact that the spin-wave excitations remain gapped
at all wave vectors corroborates that the ground state is indeed
given by our reference state \eqref{fm} and not by an in-plane
($XY$-type) ferromagnetic state.

Interestingly, since a phase with ferromagnetic long-range order sets
in, time-reversal symmetry is spontaneously broken. 
As shown in Ref.~\onlinecite{neupert12}, the
ferromagnet ground state also displays an IQHE with the Hall conductivity 
$\sigma_{xy} = e^2/h$.

%%%%%%%%%%%%%%%%%%%%%%%%%%%%%%%%%%%%%%%%%%%%%%%%%%%%%%%%%%%%%%%%%%%%%%%%%%%%%%%%%%%%%
\section{Discussion}
\label{sec:discussion}

Although we have focused our discussion on the square lattice $\pi$-flux
model, the bosonization formalisms for flat-band Chern and topological
insulators, respectively introduced in Secs.~\ref{sec:boso} and
\ref{sec:boso-sqhe}, are rather general. In principle, they can be
employed to study the flat-band ferromagnetic phase of a topological
Hubbard model whose single-particle term assumes the
$4\times 4$ matrix form \eqref{ham-0-k}, 
such as the Kane-Mele-Hubbard model 
without the Rashba spin-orbit coupling.\cite{assaad13}  
In this case, spin is a good quantum number. 
Once the coefficients $B_{i,\bk}$ [Eq.~\eqref{b-coef}] of the model
are identified, the corresponding effective (interacting) boson model
is easily determined since the coefficients \eqref{ep-b-model} and
\eqref{int-b-model} of the boson model are written in
terms of the $\hat{\bB}_{i,\bk} = B_{i,\bk} / |\bB_\bk|$ functions
(see Appendices~\ref{ap:boso} and \ref{ap:boso-spin}).  
One important point is to verify whether the condition \eqref{approx-comut} 
holds, i.e., if it is possible to define two sets of independent boson operators.       
It would be interesting to see whether the bosonization scheme could be
extended to the following cases: 
(i) Four band models where spin is not conserved. It would allow us to
consider, e. g., 
the Kane-Mele-Hubbard model with a Rashba coupling.\cite{assaad13} 
(ii) Six band models where the single-particle term assumes the form
\eqref{ham-0-k} but with $h^\sigma_\bk$ being a $3\times 3$
matrix. Two examples are the tight-binding model on the kagome
lattice\cite{tang11} and the three-orbital square lattice model.\cite{sun11}

One interesting feature of the bosonization scheme developed here is
that it allows us to analytically determine the spin-wave excitation 
spectrum of a flat-band ferromagnet with topologically nontrivial
single-particle bands. At the moment, it is not clear
how to compare the results derived in Secs.~\ref{sec:boso-model} and
\ref{sec:boso-model02}  with other approximation schemes.
For the 2DEG at filling factor $\nu=1$, the
noninteracting term of the effective boson model derived within the
bosonization formalism\cite{doretto05} describes the magnetic exciton
excitations of the quantum Hall ferromagnetic ground state. The energy of
the bosons are {\it exactly} equal to the magnetic exciton dispersion
relation derived by Kallin and Halperin\cite{kallin84} within
diagrammatic calculations. Such a diagrammatic formalism is indeed
equivalent to the so-called 
time-dependent Hartree-Fock approximation.\cite{macdonald85}  
Although for flat-band Chern and topological insulators only an 
approximated bosonic
representation for the electron density operator can be derived
[Eq.~\eqref{rho-boso02}], we expected that, due to the analogy with
the 2DEG at $\nu=1$ (see discussion at the end of
Sec.~\ref{sec:boso}),  the spin-wave dispersion relations determined  
in Secs.~\ref{sec:boso-model} and \ref{sec:boso-model02}  
agree with a time-dependent Hartree-Fock analysis of the
topological Hubbard models \eqref{ham} and \eqref{ham-z2}.

A second interesting aspect of the bosonization scheme for flat-band
Chern and topological insulators is that it provides an interaction
between the bosons (spin-waves).
A detailed study of the consequences of the spin-wave--spin-wave coupling
is beyond the scope of the present paper. 
It would be interesting, for instance, to verify whether the
boson-boson interaction \eqref{h-boso4} yields two-spin-wave bound states
and whether such bound states are related to possible topological
excitations of the flat-band ferromagnetic state \eqref{fm}. 
Such an analysis is motivated by the fact that for the 2DEG at 
filling factor $\nu=1$, the boson-boson coupling derived within the
bosonization formalism\cite{doretto05} gives rise to two-boson 
bound states that are related to skyrmion-antiskyrmion pair
excitations (the charged excitation of the 2DEG at $\nu=1$ is
described as a topological excitation, quantum Hall skyrmion, of the
quantum Hall ferromagnetic ground state).      
This set of results allows us to properly treat the skyrmion as an electron
bound to a certain number of boson (spin-waves).\cite{doretto05-1} 
We defer the analysis of the effects of the boson-boson interacting
\eqref{h-boso4} to a future publication.

%%%%%%%%%%%%%%%%%%%%%%%%%%%%%%%%%%%%%%%%%%%%%%%%%%%%%%%%%%%%%%%%%%%%%%%%%%%%%%%%%%%%%
\section{Summary}
\label{sec:summary}

In this paper, we have considered the flat-band ferromagnetic phase of 
a correlated Chern insulator and a correlated Z$_2$ topological
insulator and analytically calculated the corresponding spin-wave
excitation spectra. In particular, we have considered two variants of a
topological Hubbard model, namely, one with broken time-reversal symmetry 
(Chern insulator) and another one invariant under time-reversal
symmetry (topological insulator). In both cases, the
single-particle term is the square lattice $\pi$-flux model
with topologically nontrivial and nearly flat bands. 
The spin-wave dispersion relation has been determined within a bosonization
scheme similar to the formalism\cite{doretto05} previously developed
for the 2DEG at filling factor $\nu=1$. Here, we showed that the 
formalism\cite{doretto05} can indeed be generalized for flat-band
Chern and topological insulators. 

The investigation of the spin-wave excitation spectrum indicates the
stability of the flat-band ferromagnetic phase. 
Generically, we obtain two spin-wave excitation branches due to the 
two-atom basis of the lattice model. 
We find that the correlated flat-band Chern insulator has a gapless
excitation spectrum: the Goldstone mode is associated with a
spontaneous SU(2) spin-symmetry breaking.  
For the correlated flat-band topological insulator with preserved time-reversal symmetry, 
the excitation spectrum is gapped since the flat-band ferromagnetic ground state preserves the
U(1) spin rotation symmetry of the Hamiltonian.
Moreover, within the bosonization scheme,
we find a spin-wave-spin-wave interaction. Due to the analogies with
the 2DEG at filling factor $\nu=1$, we expect that such coupling may
give rise to two-spin-wave bound states.

%%%%%%%%%%%%%%%%%%%%%%%%%%%%%%%%%%%%%%%%%%%%%%%%%%%%%%%%%%%%%%%%%%%%%%%%%%%%%%%%%%%%%
\acknowledgments

R.L.D. kindly acknowledges Faepex/PAPDIC and 
FAPESP, Project No.~2010/00479-6, for the financial support.

%%%%%%%%%%%%%%%%%%%%%%%%%%%%%%%%%%%%%%%%%%%%%%%%%%%%%%%%%%%%%%%%%%%%%%%%%%%%%%%%%%%%
\appendix

\section{Symmetries of the $\pi$-flux model: time-reversal}
\label{ap:symmetries}

In this section, we discuss the behavior of the noninteracting
fermion model \eqref{ham-0} under time-reversal. 

The time-reversal operator $\mathcal{T}$ is defined as
\begin{equation}
 \mathcal{T} = i\left( \sigma_y \otimes I \right)K,
\end{equation}
where $K$ denotes complex conjugation and $I$ is the $2\times 2$ 
identity matrix.
Invariance under time reversal, i.e.,  $[\mathcal{H}_0,\mathcal{T}] = 0$, 
implies that\cite{fu07}
\[
  \mathcal{T}\mathcal{H}_\bk\mathcal{T}^{-1} = \mathcal{H}_{-\bk},
\]
where $\mathcal{H}_\bk$ is the matrix \eqref{hk}. Since
\[
   \mathcal{T}\mathcal{H}_\bk\mathcal{T}^{-1} = 
   \left( \begin{array}{cc} 
               h^{\downarrow\, *}_\bk  &  0 \\
                              0               &  h^{\uparrow\,*}_\bk 
           \end{array} \right),
\]
invariance under time-reversal
implies that $h^\downarrow_\bk = h^{\uparrow\,*}_{-\bk}$ as mentioned
in Sec.~\ref{sec:TBmodel}.

Alternatively, we can follow Lu and Kane\cite{fu07} and write the
matrix $\mathcal{H}_\bk$ in terms of the five $4 \times 4$ Dirac
matrices 
\[
 \Gamma^{1,2,3,4,5} = \left( I \,\otimes\, \tau_x, 
                                           I \,\otimes \,\tau_y, 
                             \sigma_x\, \otimes\, \tau_z, 
                             \sigma_y\, \otimes\, \tau_z, 
                             \sigma_z\, \otimes\, \tau_z \right), 
\]
and their ten commutators
$
  \Gamma^{ij} = [\Gamma^i,\Gamma^j]/(2i).
$
Here, $\sigma_{x,y,z}$ and $\tau_{x,y,z}$ are $2\times 2$ Pauli
matrices respectively related to spin and sublattice.
The Dirac matrices obey the Clifford algebra 
$
   \Gamma^i\Gamma^j + \Gamma^i\Gamma^j = 2\delta_{ij}I.
$
For the Chern insulator discussed in Secs.~\ref{sec:TBmodel} and \ref{sec:model0}, 
\begin{equation}
   \mathcal{H}_\bk = B_{1,\bk}\Gamma^1 + B_{2,\bk}\Gamma^2 + B_{3,\bk}\Gamma^{12} 
\label{aux-time1}
\end{equation}
while, for the topological insulator considered in
Secs.~\ref{sec:TBmodel} and \ref{sec:model02},
\begin{equation}
   \mathcal{H}_\bk = B_{1,\bk}\Gamma^1 + B_{2,\bk}\Gamma^{51} + B_{3,\bk}\Gamma^{12}. 
\label{aux-time2}
\end{equation}
Since $B_{i,\bk} = B_{i,-\bk}$,
\[
   \mathcal{T}\Gamma^i\mathcal{T}^{-1} = \left\{
     \begin{array}{cl}
       +\Gamma^1, & i=1, \\
       -\Gamma^i, & i=2,3,4,5,
     \end{array} \right.
\]
$\mathcal{T}\Gamma^{12}\mathcal{T}^{-1} = \Gamma^{12}$,
and $ \mathcal{T}\Gamma^{51}\mathcal{T}^{-1} = \Gamma^{51}$,
we see that only the Hamiltonian \eqref{aux-time2} is invariant under
time-reversal.

%%%%%%%%%%%%%%%%%%%%%%%%%%%%%%%%%%%%%%%%%%%%%%%%%%%%%%%%%%%%%%%%%%%%%%%%%%%%%%%%%%%%
\section{Details of the bosonization scheme for flat-band Chern insulators}
\label{ap:boso}

In this section, we quote the expressions in terms of the coefficients
$B_{i,\bk}$ of some quantities related to the bosonization scheme
discussed in Sec.~\ref{sec:hall}.  

Let us first consider the diagonalization of the noninteracting
Hamiltonian \eqref{ham-0-k}.
It is useful to write the coefficients $u_\bk$ and $v_\bk$ of the
Bogoliubov transformation \eqref{bogo-qhe} as
\begin{eqnarray}
 u_\bk &=& \exp(+i\phi_\bk/2)\cos(\theta_\bk/2),
\nonumber \\
 v_\bk &=& \exp(-i\phi_\bk/2)\sin(\theta_\bk/2).
\label{coef-bogo} 
\end{eqnarray}
Due to the form of the matrix $h^\uparrow_\bk$ [see Eq.~\eqref{hkup}], it is
interesting to introduce the following relations between the  
$\phi_\bk$ and $\theta_\bk$ functions and the coefficients $B_{i,\bk}$  
[Eq.~\eqref{b-coef}]:  
\begin{eqnarray}
 \bhat_{1,\bk} &=& \sin\theta_\bk\cos\phi_\bk,
 \;\;\;\;\;\;\;\;\;\;\;
 \bhat_{2,\bk} = \sin\theta_\bk\sin\phi_\bk,
\nonumber \\
 \bhat_{3,\bk} &=& \cos\theta_\bk,
\nonumber \\
 \sin\theta_\bk &=& \sqrt{\bhat^2_{1,\bk} + \bhat^2_{2,\bk}},
 \;\;\;\;\;
 \tan\phi_\bk = \frac{\bhat_{2,\bk}}{\bhat_{1,\bk}},
\label{phi-theta}
\end{eqnarray} 
where $\hat{\bB}_\bk \equiv \bB_\bk / |\bB_\bk|$.
It then follows that
\begin{eqnarray}
               |u_\bk|^2 &=& \frac{1}{2}\left( 1 + \bhat_{3,\bk} \right), 
\;\;\;\;\;\;\;
                  |v_\bk|^2 = \frac{1}{2}\left( 1 - \bhat_{3,\bk} \right),  
\nonumber \\
u^*_\bk v_\bk &=& \frac{1}{2}\left( \bhat_{1,\bk} +i\bhat_{2,\bk} \right).  
\label{coef-bogo2}
\end{eqnarray}
With the aid of Eqs.~\eqref{coef-bogo}-\eqref{coef-bogo2}, it is easy
to show that the Hamiltonian \eqref{diag-ham-0} is the diagonal form
of \eqref{ham-0-k}.

Concerning the algebra of the projected spin and electron density
operators, in addition to the commutator \eqref{comut-ss}, it follows
from Eqs.~\eqref{spin-alpha-pm-op} and \eqref{proj-dens-op} that
\begin{widetext}
\begin{eqnarray}
  [ \bar{\rho}_{a\,\sigma}(\bk), \bar{S}^+_{\bq,\alpha}  ] &=&
   \sum_\bp  \left[  \delta_{\sigma,\uparrow}G_a(\bp-\bq,\bk)g_\alpha(\bp,\bq)
                            -\delta_{\sigma,\downarrow}G_a(\bp,\bk)g_\alpha(\bp-\bk,\bq)  \right]
                    c^\dagger_{\bp-\bq-\bk\,\uparrow}c_{\bp\,\downarrow},
\label{comut-s+rho} \\
\,   [ \bar{\rho}_{a\,\sigma}(\bk), \bar{S}^-_{\bq,\alpha}  ] &=&
       \sum_\bp  \left[  \delta_{\sigma,\downarrow}G_a(\bp-\bq,\bk)g_\alpha(\bp,\bq)
                            -\delta_{\sigma,\uparrow}G_a(\bp,\bk)g_\alpha(\bp-\bk,\bq)  \right]
                    c^\dagger_{\bp-\bq-\bk\,\downarrow}c_{\bp\,\uparrow},
\label{comut-s-rho} \\
\, [ \bar{\rho}_{a\,\sigma}(\bk), \bar{\rho}_{b\,\sigma'}(\bq) ] &=& \delta_{\sigma,\sigma'} 
   \sum_\bp  \left[  G_a(\bp-\bq,\bk)G_b(\bp,\bq)
                            - G_a(\bp,\bk)G_b(\bp-\bk,\bq)  \right]
                    c^\dagger_{\bp-\bq-\bk\,\sigma}c_{\bp\,\sigma},
\label{comut-rhorho}
\end{eqnarray}
with $a,b = A,B$, $\alpha=0,1$,  and the $G_a(\bp,\bq)$ and
$g_\alpha(\bp,\bq)$ functions respectively given by  
Eqs.~\eqref{gab-functions} and \eqref{g-func-hall}.

Within the approximation \eqref{assumption}, the commutator
\eqref{comut-ss} assumes the form \eqref{approx-comut}.  
With the aid of Eq.~\eqref{coef-bogo2}, it is possible to show that 
($\alpha \not= \beta$)
\begin{equation}
 \sum_\bp g_\alpha(\bp,\bq) g_\beta(\bp-\bq,-\bq)
     = \frac{1}{2}\sum_\bp \left[ (-1)\left( \hat{B}_{3,\bp} + \hat{B}_{3,\bp-\bq} \right)
     +  i(-1)^\alpha\left( \hat{B}_{1,\bp-\bq}\hat{B}_{2,\bp}  - \hat{B}_{2,\bp-\bq}\hat{B}_{1,\bp}  \right)
        \right].
\label{f-alpha-beta}
\end{equation}
Since for the $\pi$-flux model, $B_{i,\bk} = B_{i,-\bk}$ [see Eq.~\eqref{b-coef}]   
it is then easy to show that Eq.~\eqref{f-alpha-beta} vanishes.

Similar to Eq.~\eqref{f-alpha-beta}, it is possible to write the $F_{\alpha,\bq}$
function in terms of the coefficients $B_{i,\bk}$. Using
Eq.~\eqref{coef-bogo2}, we show that
\begin{eqnarray}
 F^2_{\alpha,\bq}  = \frac{1}{2}\sum_\bp \left[ 1 + \hat{B}_{3,\bp}\hat{B}_{3,\bp-\bq}
                             (-1)^\alpha\left( \hat{B}_{1,\bp}\hat{B}_{1,\bp-\bq}
                               + \hat{B}_{2,\bp}\hat{B}_{2,\bp-\bq}  \right)
                \right].
\label{f-func-hall}
\end{eqnarray}

The (approximated) boson representation of the projected electron-density operator is given by Eq.~\eqref{rho-boso02} 
where the $\gm_{\alpha\,\beta\,a\,\alpha}(\bk,\bq)$ function is 
defined as
\begin{eqnarray}
 \gm_{\alpha\,\beta\,a\,\uparrow}(\bk,\bq) &=& 
        -\frac{1}{F_{\alpha,\bq} F_{\beta,\bk+\bq}}\sum_\bp
         G_a(\bp,\bk)g_\alpha(\bp-\bk,\bq)g_\beta(\bp-\bk-\bq,-\bk-\bq),
\nonumber \\
\label{galphabeta-hall01}&& \\
 \gm_{\alpha\,\beta\,a\,\downarrow}(\bk,\bq) &=& 
        +\frac{1}{F_{\alpha,\bq} F_{\beta,\bk+\bq}}\sum_\bp
          G_a(\bp-\bq,\bk)g_\alpha(\bp,\bq)g_\beta(\bp-\bk-\bq,-\bk-\bq).
\nonumber 
\end{eqnarray}
Again, using Eqs.~\eqref{coef-bogo2}, we find after some algebra that 
\begin{eqnarray}
 && \gm_{\alpha\,\beta\,a\,\uparrow}(\bk,\bq) = 
     -\frac{1}{8}[\delta_{a,A} + \delta_{a,B}(-1)^{\alpha+\beta}]
      \frac{1}{F_{\alpha,\bq} F_{\beta,\bk+\bq}}\sum_\bp
      1 - 3(-1)^a\bhat_{3,\bp} 
\nonumber \\
 && \nonumber \\
 &&\; 
     + \bhat_{3,\bp-\bq}\bhat_{3,\bp+\bk}
     + \bhat_{3,\bp-\bq}\bhat_{3,\bp}
     + \bhat_{3,\bp+\bk}\bhat_{3,\bp}
     - (-1)^a\bhat_{3,\bp-\bq}\bhat_{3,\bp}\bhat_{3,\bp+\bk}
\nonumber \\
 && \nonumber \\
 &&\; 
      + (-1)^{\alpha}[\bhat_{1,\bp-\bq}\bhat_{1,\bp} + \bhat_{2,\bp-\bq}\bhat_{2,\bp}
           + i(-1)^a(\bhat_{1,\bp-\bq}\bhat_{2,\bp} - \bhat_{2,\bp-\bq}\bhat_{1,\bp} ) ]
        [ 1 - (-1)^a\bhat_{3,\bp + \bk} ]
\nonumber \\
 && \nonumber \\
 &&\; 
      + (-1)^{\beta}[\bhat_{1,\bp-\bq}\bhat_{1,\bp+\bk} + \bhat_{2,\bp-\bq}\bhat_{2,\bp+\bk} 
           - i(-1)^a( \bhat_{1,\bp-\bq}\bhat_{2,\bp+\bk} - \bhat_{2,\bp-\bq}\bhat_{1,\bp+\bk} )]
        [ 1 - (-1)^a\bhat_{3,\bp} ]
\nonumber \\
 && \nonumber \\
 && \; 
      + (-1)^{\alpha+\beta}[\bhat_{1,\bp+\bk}\bhat_{1,\bp} + \bhat_{2,\bp+\bk}\bhat_{2,\bp}
           + i(-1)^a( \bhat_{1,\bp+\bk}\bhat_{2,\bp}  - \bhat_{2,\bp+\bk}\bhat_{1,\bp}  )]
        [ 1 + (-1)^a\bhat_{3,\bp - \bq} ].
\label{galphabeta-hall02}
\end{eqnarray}
The expression of $\gm_{\alpha\,\beta\,a\,\downarrow}(\bk,\bq)$ can be
derived from Eq.~\eqref{galphabeta-hall02} using the fact that
$\gm_{\alpha\,\beta\,a\,\downarrow}(\bk,\bq) = -\gm^*_{\alpha\,\beta\,a\,\uparrow}(-\bk,-\bq)$.

Finally, the bosonic representation of the projected single-particle
Hamiltonian \eqref{ham-0-proj}. The first step is the calculation of
the commutator 
\begin{eqnarray}
[\bar{\mathcal{H}}_0, b^\dagger_{\alpha,\bq}] &=&
     \sum_\bp \left( \omega_{c,\bp-\bq} - \omega_{c,\bp} \right) \frac{g_\alpha(\bp,\bq)}{F_{\alpha,\bq} }
     c_{\bp-\bq\,\downarrow}c_{\bp\,\uparrow} 
 = \sum_{\beta,\bk,\bp}\left( \omega_{c,\bp-\bq} - \omega_{c,\bp} \right) \frac{g_\alpha(\bp,\bq)}{F_{\alpha,\bq} } 
           H_\beta(\bk,\bp,\bq) b^\dagger_{\beta,\bk}.
\label{boso-aux1}
\end{eqnarray}
Here, the second equality follows from Eq.~\eqref{cc1}. Note that the expansion  
\begin{equation}
  \bar{\mathcal{H}}_{0,B} = \sum_{\alpha,\beta}\sum_{\bk,\bp,\bq}
             \left( \omega_{c,\bp-\bq} - \omega_{c,\bp} \right) \frac{g_\alpha(\bp,\bq)}{F_{\alpha,\bq} } 
             H_\beta(\bk,\bp,\bq) b^\dagger_{\beta,\bk}b_{\alpha,\bk}
\end{equation}
of $\bar{\mathcal{H}}_0$ in terms of the boson operators $b_\alpha$
satisfies the commutator \eqref{boso-aux1}. Keeping only the term 
$\bk = \bq$ in the momentum sum above and using
Eq.~\eqref{h-func-approx}, we have 
\begin{equation}
 \bar{\mathcal{H}}_{0,B} = E_0 +
            \sum_{\alpha,\beta}\sum_\bq\bar{\omega}_{\alpha\beta}(\bq) b^\dagger_{\beta,\bq} b_{\alpha,\bq},
\end{equation}
where $E_0 = 2.44N$ is a constant related to the action of
$\bar{\mathcal{H}}_0$ in the reference state \eqref{fm} and
\begin{equation}
  \bar{\omega}_{\alpha\beta}(\bq) = \frac{1}{F_{\alpha,\bq}F_{\beta,\bq} }\sum_\bp
                    \left( \omega_{c,\bp-\bq} - \omega_{c,\bp} \right) 
                     g_\alpha(\bp,\bq)g_\beta(\bp -\bq,-\bq).
\label{boso-aux2}
\end{equation}
In the flat-band limit, $\bar{\omega}_{\alpha\beta}(\bq) = 0$ since
$\omega_{c,\bp} = 0$. In the nearly flat-band limit considered in
Sec.~\ref{sec:boso-model},  the coefficient
$\bar{\omega}_{\alpha\beta}(\bq)$ can be nonzero. However, for the
$\pi$-flux model, it is possible to show that
$\bar{\omega}_{\alpha\beta}(\bq)$ vanishes due to the fact that 
$B_{i,\bk} = B_{i,-\bk}$.

%%%%%%%%%%%%%%%%%%%%%%%%%%%%%%%%%%%%%%%%%%%%%%%%%%%%%%%%%%%%%%%%%%%%%%%%%%%%%%%%%%%%
\section{Details of the bosonization scheme for flat-band topological insulators}
\label{ap:boso-spin}

Similar to the previous section, we here provide the expansion in
terms of the coefficients $B_{i,\bk}$ of some functions related to the
bosonization formalism introduced in Sec.~\ref{sec:shall}. 

The $F_{\alpha,\bq}$ function \eqref{f-func-sqhe} reads
\begin{eqnarray}
 F^2_{\alpha,\bq} =  \frac{1}{2}\sum_\bp \left[  (-1)^\alpha (1 - \hat{B}_{3,\bp}\hat{B}_{3,-\bp+\bq})
                       +    \hat{B}_{1,\bp}\hat{B}_{1,-\bp+\bq} + \hat{B}_{2,\bp}\hat{B}_{2,-\bp+\bq}  
                \right].
\label{f-func-sqhe2}
\end{eqnarray}

The algebra of the projected spin and electron-density operators. From
Eqs.~\eqref{spin-alpha-pm-op2} and \eqref{density-op-proj2}, we show
that, in addition to the commutator \eqref{comut-ss-z2}, the following
commutation relations hold
\begin{eqnarray}
  [ \bar{\rho}_{a\,\sigma}(\bk), \bar{S}^+_{\bq,\alpha}  ] &=&
   \sum_\bp  \left[  \delta_{\sigma,\uparrow}G_{a\,\sigma}(\bp-\bq,\bk)g^*_\alpha(-\bp,-\bq)
                            -\delta_{\sigma,\downarrow}G_{a\,\sigma}(\bp,\bk)g^*_\alpha(-\bp+\bk,-\bq)  \right]
                    c^\dagger_{\bp-\bq-\bk\,\uparrow}c_{\bp\,\downarrow},
\\
\,   [ \bar{\rho}_{a\,\sigma}(\bk), \bar{S}^-_{\bq,\alpha}  ] &=&
       \sum_\bp  \left[  \delta_{\sigma,\downarrow}G_{a\,\sigma}(\bp-\bq,\bk)g_\alpha(\bp,\bq)
                            -\delta_{\sigma,\uparrow}G_{a\,\sigma}(\bp,\bk)g_\alpha(\bp-\bk,\bq)  \right]
                    c^\dagger_{\bp-\bq-\bk\,\downarrow}c_{\bp\,\uparrow},
\\
\, [ \bar{\rho}_{a\,\sigma}(\bk), \bar{\rho}_{b\,\sigma'}(\bq) ] &=& \delta_{\sigma,\sigma'} 
   \sum_\bp  \left[  G_{a\,\sigma}(\bp-\bq,\bk)G_{b\,\sigma'}(\bp,\bq)
                            - G_{a\,\sigma}(\bp,\bk)G_{b\,\sigma'}(\bp-\bk,\bq)  \right]
                    c^\dagger_{\bp-\bq-\bk\,\sigma}c_{\bp\,\sigma},
\end{eqnarray}
with $a,b = A,B$, $\alpha=0,1$,  and the $g_\alpha(\bp,\bq)$ 
and $G_{a\,\sigma}(\bp,\bq)$ functions respectively given by  
Eqs.~\eqref{g-func-shall} and \eqref{gabsigma-functions}.

The $\gm_{\alpha\,\beta\,a\,\sigma}(\bk,\bq)$ function, which is related to the
bosonic representation of the projected electron-density operator
\eqref{density-op-proj2}, is defined as
\begin{eqnarray}
 \gm_{\alpha\,\beta\,a\,\uparrow}(\bk,\bq) &=& 
        -\frac{1}{F_{\alpha,\bq} F_{\beta,\bk+\bq}}\sum_\bp
         G_{a\,\uparrow}(\bp,\bk)g_\alpha(\bp-\bk,\bq)g^*_\beta(-\bp+\bk+\bq,\bk+\bq),
\nonumber \\
&& \label{galphabeta-shall01}\\
 \gm_{\alpha\,\beta\,a\,\downarrow}(\bk,\bq) &=& 
        +\frac{1}{F_{\alpha,\bq} F_{\beta,\bk+\bq}}\sum_\bp
          G_{a\,\downarrow}(\bp-\bq,\bk)g_\alpha(\bp,\bq)g^*_\beta(-\bp+\bk+\bq,\bk+\bq)
\nonumber 
\end{eqnarray}
and, in terms of the coefficients $B_{i,\bq}$, it reads 
\begin{eqnarray}
 && \gm_{\alpha\,\beta\,a\,\uparrow}(\bk,\bq) = 
     -\frac{1}{8}[\delta_{a,A} + \delta_{a,B}(-1)^{\alpha+\beta}]
      \frac{1}{F_{\alpha,\bq} F_{\beta,\bk+\bq}}\sum_\bp
      (-1)^{\alpha} [1 - (-1)^a\bhat_{3,\bp} 
\nonumber \\
 && \nonumber \\
 &&\; 
     - \bhat_{3,-\bp+\bq}\bhat_{3,\bp+\bk}
     - \bhat_{3,-\bp+\bq}\bhat_{3,\bp}
     + \bhat_{3,\bp+\bk}\bhat_{3,\bp}
     - (-1)^a\bhat_{3,-\bp+\bq}\bhat_{3,\bp}\bhat_{3,\bp+\bk} ]
\nonumber \\
 && \nonumber \\
 &&\; 
      + [\bhat_{1,-\bp+\bq}\bhat_{1,\bp} + \bhat_{2,-\bp+\bq}\bhat_{2,\bp}
           + i(-1)^a(\bhat_{1,-\bp+\bq}\bhat_{2,\bp} - \bhat_{2,-\bp+\bq}\bhat_{1,\bp} ) ]
        [ 1 - (-1)^a\bhat_{3,\bp + \bk} ]
\nonumber \\
 && \nonumber \\
 &&\; 
      + (-1)^{\alpha+\beta}[\bhat_{1,-\bp+\bq}\bhat_{1,\bp+\bk} + \bhat_{2,-\bp+\bq}\bhat_{2,\bp+\bk} 
           - i(-1)^a( \bhat_{1,-\bp+\bq}\bhat_{2,\bp+\bk} - \bhat_{2,-\bp+\bq}\bhat_{1,\bp+\bk} )]
        [ 1 - (-1)^a\bhat_{3,\bp} ]
\nonumber \\
 && \nonumber \\
 && \; 
      + (-1)^{\beta}[\bhat_{1,\bp+\bk}\bhat_{1,\bp} + \bhat_{2,\bp+\bk}\bhat_{2,\bp}
           + i(-1)^a( \bhat_{1,\bp+\bk}\bhat_{2,\bp}  - \bhat_{2,\bp+\bk}\bhat_{1,\bp}  )]
        [ 1 - (-1)^a\bhat_{3,-\bp + \bq} ]
\label{galphabeta-shall02}
\end{eqnarray}
The expansion for $\gm_{\alpha\,\beta\,a\,\downarrow}(\bk,\bq)$ 
follows from Eq.~\eqref{galphabeta-shall02} using the relation
$\gm_{\alpha\,\beta\,a\,\downarrow}(\bk,\bq) = -(-1)^{\alpha+\beta}\gm_{\alpha\,\beta\,a\,\uparrow}(\bk,\bq)$.

\end{widetext}

%%%%%%%%%%%%%%%%%%%%%%%%%%%%%%%%%%%%%%%%%%%%%%%%%%%%%%%%%%%%%%%%%%%%%%%%%%%%%%%%%%%%

\end{document}